\documentclass[a4paper,fleqn,usenatbib]{mnras}
\usepackage{newtxtext,newtxmath}
\usepackage[T1]{fontenc}
\usepackage{graphicx}	
\usepackage{amsmath}	
\usepackage{enumitem}
\usepackage{rotating}
\usepackage{pdflscape}

\newcommand{\te}{\emph{TESS}}
\newcommand{\kms}{km\,s$^{-1}$}

\title[Binarity and fast rotation]{Extreme mass ratios and fast rotation in three massive binaries\thanks{Based on data obtained with \te, ESO, Espadons, TIGRE, IUE, FUSE }}

\author[Y. Naz\'e et al.]{Ya\"el~Naz\'e$^1$\thanks{F.R.S.-FNRS Senior Research Associate, email: ynaze@uliege.be}, Nikolay Britavskiy$^1$, Gregor Rauw$^1$, Jonathan Labadie-Bartz$^2$, S.~Sim\'on-D{\'\i}az$^{3,4}$ 
\\
$^1$ Groupe d'Astrophysique des Hautes Energies, STAR, Universit\'e de Li\`ege, Quartier Agora (B5c, Institut d'Astrophysique et de G\'eophysique), \\
All\'ee du 6 Ao\^ut 19c, B-4000 Sart Tilman, Li\`ege, Belgium\\
$^2$ LESIA, Paris Observatory, PSL University, CNRS, Sorbonne University, Universit\'e Paris Cit\'e, 5 place Jules Janssen, 92195 Meudon, France\\
$^3$ Instituto de Astrof\'isica de Canarias. E-38200 La Laguna, Tenerife, Spain.\\
$^4$ Departamento de Astrof\'isica, Universidad de La Laguna. E-38205 La Laguna, Tenerife, Spain.\\
}

\pubyear{2023}

\begin{document}
\label{firstpage}
\pagerange{\pageref{firstpage}--\pageref{lastpage}}
\maketitle

\begin{abstract}
The origin of rapid rotation in massive stars remains debated, although binary interactions are now often advocated as a cause. However, the broad and shallow lines in the spectra of fast rotators make direct detection of binarity difficult. In this paper, we report on the discovery and analysis of multiplicity for three fast-rotating massive stars: HD\,25631 (B3V), HD\,191495 (B0V), and HD\,46485 (O7V). They display strikingly similar \te\ light curves, with two narrow eclipses superimposed on a sinusoidal variation due to reflection effects. We complement these photometric data by spectroscopy from various instruments (X-Shooter, Espadons, FUSE...), to further constrain the nature of these systems. The detailed analyses of these data demonstrates that the companions of the massive OB stars have low masses ($\sim1$\,M$_{\odot}$) with rather large radii (2--4\,R$_{\odot}$) and low temperatures ($<15$\,kK). These companions display no UV signature, which would exclude a hot subdwarf nature, but disentangling of the large set of X-Shooter spectra of HD\,25631 revealed the typical signature of chromospheric activity in the companion's spectrum. In addition, despite the short orbital periods ($P=3-7$\,d), the fast-rotating OB-stars still display non-synchronized rotation and all systems appear young ($<$20\,Myr). This suggests that, as in a few other cases, these massive stars are paired in those systems with non-degenerate, low-mass PMS companions, implying that fast rotation would not be a consequence of a past binary interactions in their case.
\end{abstract}


\section{Introduction}

Massive stars, born with OB spectral types, are very important feedback agents in the galaxies. This feedback is exerted through their bright emissions of ionizing light, their strong wind ouflows, as well as their final supernova explosions. All these ingredients are affected by the evolutionary paths followed by these stars, which in turn depend on mass, mass-loss, metallicity but also rotation. In this context, massive stars are known to rotate faster than their low-mass counterparts, on average. Over the years, numerous observational studies have indeed been performed to determine the projected rotational velocities $v \sin(i)$ of various sets of massive stars, often with the aim of deriving the actual distribution of rotation rates after deconvolution. The majority of massive stars display rotation velocities below 200\,\kms\footnote{\citet{hol22} mentioned a fraction of 75\% of Galactic O-stars in their sample having projected rotational velocities below 150\,\kms, while \citet{duf13} estimated that a quarter of their sample have projected rotational velocities below 100\,\kms. Comparing the various samples (see references in text) leads to a threshold of 200\,\kms\ below which most OB stars are found and above which most stars have been suggested to be post-interaction products \citep{dem13}.}, although values up to 610\,\kms\ have been recorded in some cases \citep{duf11,ram13} and Be stars are known to reach 88\% of the critical velocity on average \citep{fre05}. Main-sequence massive stars clearly show a broad distribution of rotational velocities (e.g. \citealt{how97}), but details regarding the distribution shape vary between studies, as they focus on different samples. \citet{hua10} found similar rotational velocities in field and cluster stars, taking age into account, and a larger proportion of fast rotators in lower-mass B stars, compared to higher-mass ones. \citet{con77} and \citet{duf13} described the overall distribution as bimodal for Galactic O-stars and LMC B-stars, respectively. \citet{ram13,ram15} mostly used a ``peak+tail'' wording, without disagreeing on a bimodality vocabulary, while \citet{abt02} derived an exponential-like distribution for early B-stars, rejecting bimodality, and \citet{hua10} have different distribution shapes depending on the mass range considered, some with a clear bimodality, some without. Finally, \citet{duf13} also found that the distribution of a previously published sample showed no bimodality, in contrast to theirs. What all authors agree on, however, is the presence of a significant fraction of fast rotating stars.

Such a fast rotation strongly impacts the stellar evolution \citep{bro11,eks12,mae15} and can affect the nature of the remnant \citep{pet05mag}. It is therefore of utmost interest to understand the origin of rapid rotation, for which several scenarios have been proposed. For example, the present-day rotation rate could be solely tied to the early formation stages. As stars are born with a range of rotational velocities, some of them would naturally display larger values than the others \citep{zor97}. Indeed, \citet{hua10} found that half of young B-stars are fast rotators displaying rotational velocities between 40 and 80\% of the critical values. Simulations by \citet{ros12} also established that massive ($M>6$\,M$_{\odot}$) stars are preferentially born fast rotating, magnetic torques being usually unable to spin-down the massive pre-main sequence stars at both main and final formation stages. Whether (and how long) pre-main sequence stars are able to retain their circumstellar disk (a factor that is influenced by the environment) also play an important role in this context \citep{bas20}. The initial rotation rate could in addition further increase during main sequence evolution \citep{eks08}. An alternative scenario, with some empirical support \citep{hol22,bri23}, relies on interactions within a binary system \citep{pod92,van97,dem13}. There are several ways in which the rotation of a star may be affected by its companion. Tides will apply torques able to change the rotation of a star, e.g. forcing corotation \citep{zah75,hut81,son13}. A transfer of mass will also lead to a transfer of angular momentum \citep{pac81,pol91,pet05,der10}. The merging of stars can also produce a star with peculiar properties, including a fast rotation (e.g. \citealt{jia13}) - although internal readjustment of the merger product and magnetic braking post-merger could quickly slow down the rotational velocity \citep{sch16,sch20}.

Observational evidence is being gathered to discriminate between these possibilities. As stellar rotation induces some internal mixing, in the early phases of evolution, it can produce a change in chemical abundances observed at the stellar surfaces. Abundance studies have, however, revealed that not all stars followed the predictions of single-star evolutionary models. \citet{hun07,hun09} unveiled a population of fast rotators without Nitrogen enrichment and of N-enriched slow rotators (the latter group being later confirmed e.g. by \citealt{gri17}), whereas a direct link between rotation and abundance was expected. Focusing on O-type fast rotators, \citet{caz17theo} further showed that single-star models could certainly not reproduce the diversity of observed chemical properties in those stars. Also, for the fast-rotating massive stars of type Be, the presence of companions now appears ubiquitous (e.g. \citealt{kle19}), and their nature was found in several cases to correspond to post-interaction stripped stars (e.g. \citealt{wan21}). Finally, the fraction of fast rotators in the stellar population \citep{dem13} as well as the fraction of runaways amongst fast rotators \citep{bou18,san22,bri23} are well reproduced by taking only the binary channel into account. The role of binary interactions is thus nowadays considered as predominant, if not absolutely required to explain the observed population of fast rotators. 

In order to shed further light on the origin of fast rotation, we are performing in-depth studies of fast-rotating OB-type stars. After a focus on fast rotators with O spectral type \citep{bri23}, this paper focuses on three cases displaying light curves with specific binarity signatures: HD\,25631 (B3V, $v \sin(i)$=221\,\kms), HD\,191495 (B0V, $v \sin(i)$=201\,\kms), and HD\,46485 (O7V, $v \sin(i)$=334\,\kms). After presenting the data used in our study (Sect. 2), we analyze all available information (Sect. 3) before discussing the results (Sect. 4) and concluding (Sect. 5).

\begin{figure*}
  \begin{center}
    \includegraphics[width=8.5cm]{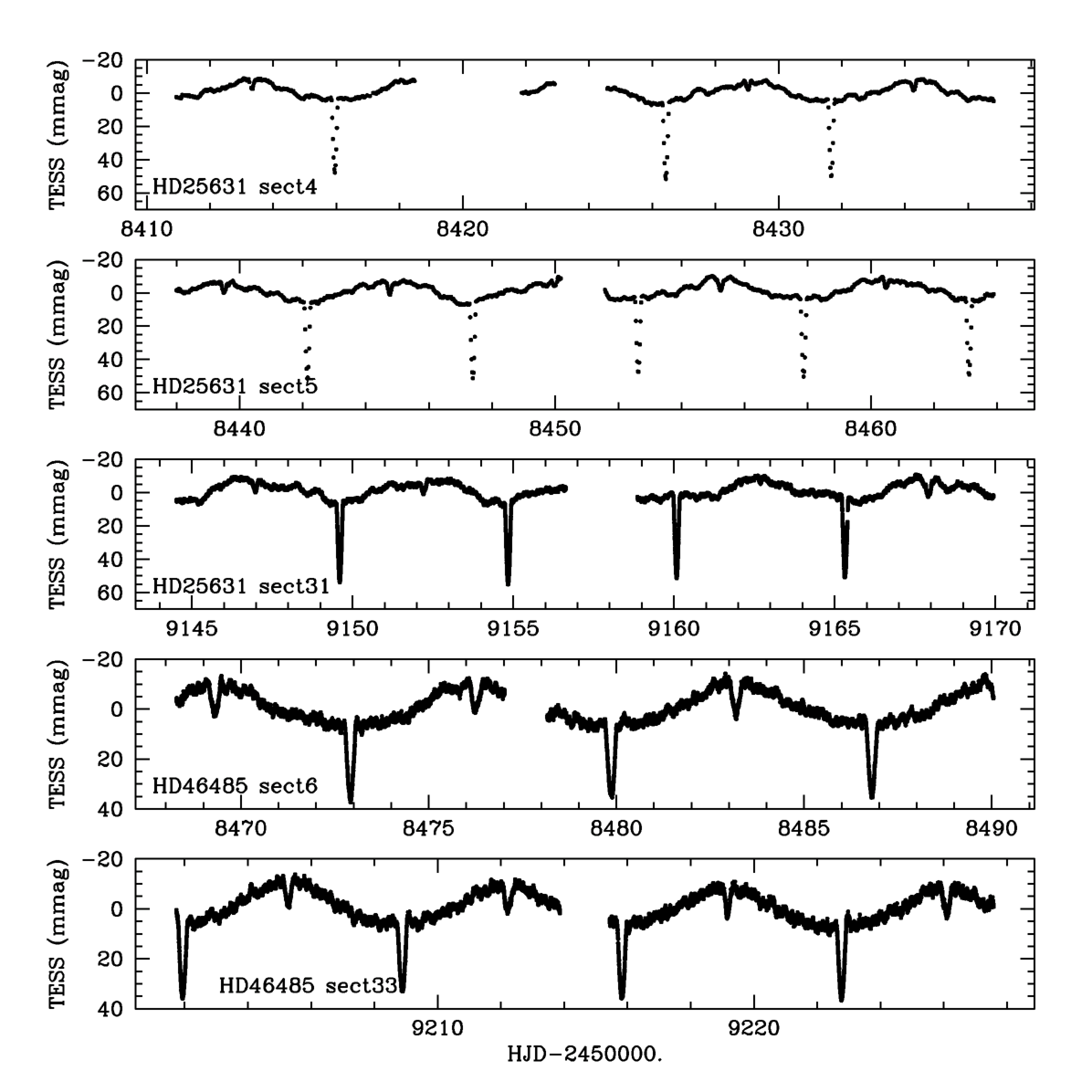}
    \includegraphics[width=8.5cm]{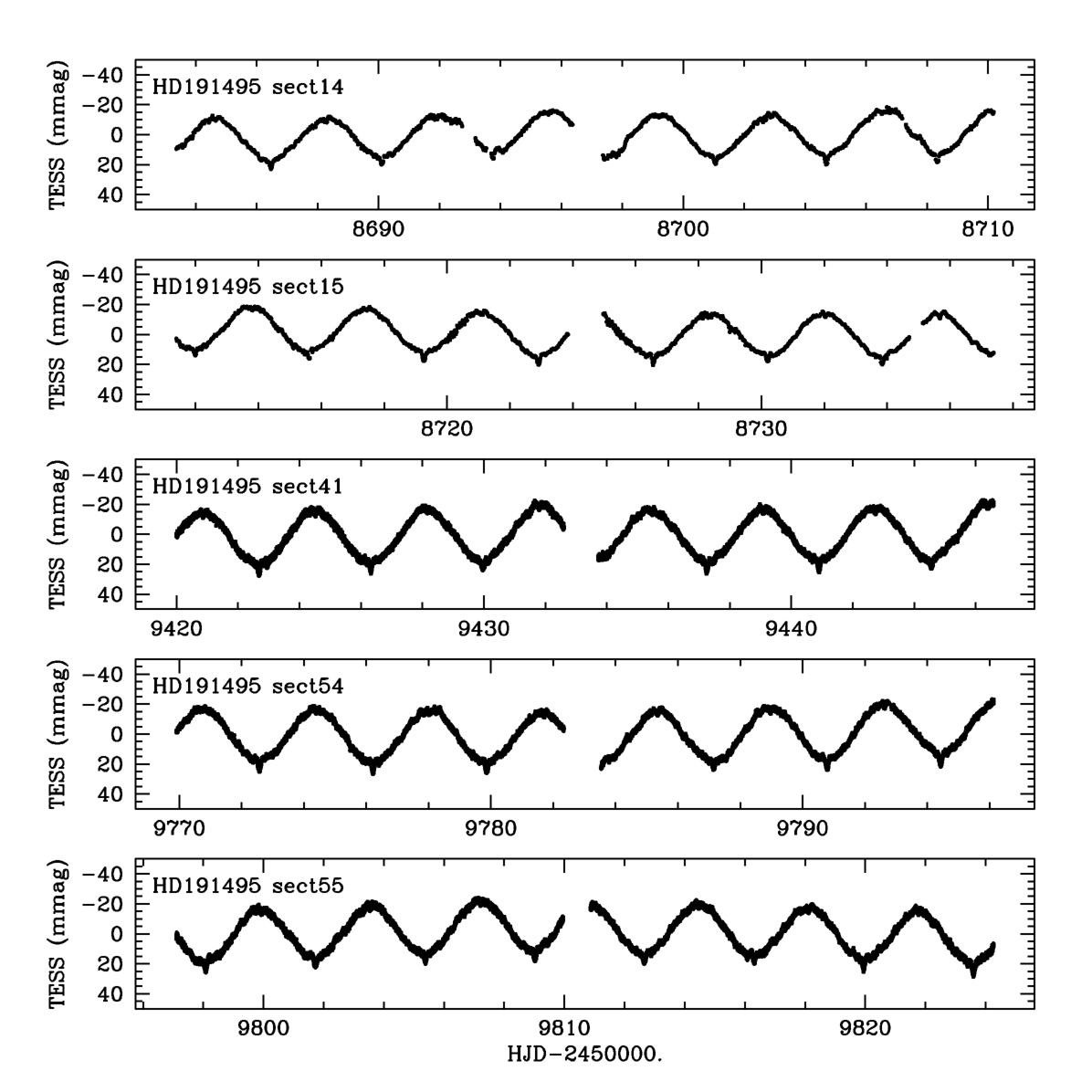}
  \end{center}
  \caption{The \te\ light curves of the three targets.}
\label{lcall}
\end{figure*}

\section{The different datasets}

\subsection{\te\ Light Curves}

Since 2018, the \te\ satellite provides high-cadence photometry for most of the sky, pointings being made in ``sectors'' of duration $\sim$25\,d \citep{ric15}. HD\,25631 was observed in Sectors 4 and 5 with the usual 30\,min cadence, but it was a pre-selected target with a more intense monitoring (2\,min cadence) in Sector 31. HD\,46485 was observed in both Sectors 6 and 33 with 2\,min cadence, while HD\,191495 was observed in Sectors 14 and 15 with 30\,min cadence and in Sectors 41, 54, and 55 with 2\,min cadence. Since \te\ pixels are large (21\arcsec), crowding may be an issue. However, we note that there are no bright ($\Delta G < 2.5$\,mag) and close (within 1\arcmin) neighbours to these three targets in the Gaia-DR3 catalog. Therefore, the light curves appear representative of the massive objects themselves.

For \te\ data, the main reduction steps (pixel-level calibration, background subtraction, flat-fielding, and bias subtraction) are done by a pipeline similar to that designed for the {\it Kepler} mission. All reduced data are then available from the public MAST archive portal\footnote{https://mast.stsci.edu/portal/Mashup/Clients/Mast/Portal.html}.

The 2\,min-cadence time-series are further corrected for crowding, the limited size of the aperture, and instrumental systematics. We nevertheless checked the absence of problems by comparing these conditioned light curves (so-called ``PDCSAP'') to the Simple Aperture Photometry (``SAP'') light curve. In addition, only the best quality (quality flag=0) data points were kept for our analyses. 

The 30\,min-cadence time-series were extracted from \te\ public data with the Python package Lightkurve\footnote{https://docs.lightkurve.org/}. Lightkurve allowed us to perform aperture photometry on 50px$\times$50px cutouts of \te\ full frame images. The source masks were defined as pixels with fluxes above some threshold (50 and 20 Median Absolute Deviation over the median flux for HD\,25631 and HD\,191495, respectively ). To avoid neighbouring sources, the background mask was defined by pixels with fluxes below the median flux (i.e. below the null threshold). Several background correction methods were tried: a simple median of the background pixels as well as a principal component analysis (PCA) with two or five components. For HD\,191495, the PCA trials yielded noisier light curves, so median subtraction was favored, while for HD\,25631, the traditional PCA-5 method yielded excellent results and was thus adopted. For Sector 4, data points with Julian dates 2\,458\,420--1.85 were taken out as they displayed a larger mean than other data. Finally, all data points with errors larger than the mean of the errors plus three times their 1$\sigma$ dispersion were also discarded. 

Whatever the cadence, the raw fluxes were converted into magnitudes using $mag=-2.5\times \log(flux)$ and their mean was then subtracted. For HD\,191495, the two parts of the Sector 41 light curve had different mean levels, so that averages were calculated for each part. As the targets were observed over several sectors, the final light curves were combined. All light curves are shown in Fig. \ref{lcall}.

\begin{table}
  \scriptsize
  \caption{Radial velocities derived from X-Shooter data of HD\,25631. 1-$\sigma$ error bars amount to 0.5\,\kms. \label{rv25631}}
  \begin{tabular}{lc|lc|lc}
    \hline
$HJD$ & $RV$& $HJD$ & $RV$ & $HJD$ & $RV$\\
$-2.45e6$& (\kms) & $-2.45e6$& (\kms) & $-2.45e6$& (\kms) \\
\hline                                               
5170.5684 & 45.5 &5891.7417 & -7.5 &  6304.5183 &  1.5  \\ 
5186.6418 & 43.5 &5907.6289 & -4.0 &  6305.5313 &-14.0  \\ 
5250.5771 & 21.0 &5913.7587 & 26.5 &  6306.5126 &  4.5  \\
5512.7636 &  6.5 &5976.5806 & 26.0 &  6370.4901 & 40.0  \\
5516.6045 & 44.5 &5985.5806 & -4.0 &  6526.8728 & 18.0  \\
5517.6250 & 25.5 &5992.5737 & 38.0 &  6579.6510 & 34.0  \\
5528.7654 & 16.5 &6143.8283 &  7.0 &  6583.6696 &-12.5  \\
5529.7617 &-13.5 &6151.9024 & 23.0 &  6644.5775 & 15.5  \\
5557.7053 & 30.5 &6166.8738 & 39.5 &  6693.6443 &-11.5  \\
5559.7191 & 21.0 &6167.8654 & 17.0 &  6889.8627 & 50.5  \\
5561.6685 &-17.5 &6187.8236 & 41.5 &  6960.6611 & -9.5  \\
5563.6875 & 37.0 &6189.8186 &-14.0 &  6990.6439 & 14.5  \\
5565.6907 & -3.0 &6190.8125 & -5.5 &  6999.5709 & 45.0  \\
5566.6922 & -6.0 &6201.8098 & 15.5 &  7246.8532 & 31.5  \\
5567.6335 & 14.5 &6209.7717 & 13.5 &  7278.8971 & 14.0  \\
5568.6623 & 44.0 &6210.7728 &-12.5 &  7287.7528 & 48.5  \\
5569.6378 & 45.0 &6211.7313 &-10.0 &  7361.5754 & 46.5  \\
5570.6598 &  7.5 &6215.7170 & -6.5 &  7363.7762 & -4.5  \\
5631.5565 & 38.5 &6219.6669 & 30.5 &  7371.6932 & 46.0  \\
5791.8338 &-11.5 &6245.6762 & 37.0 &  7630.8194 & -9.0  \\
5799.8402 & 38.0 &6254.6539 & 28.0 &  7666.7370 & 10.0  \\
5803.9184 & 31.0 &6274.7741 & 10.0 &  7975.8317 & 12.0  \\
5813.8828 & -8.0 &6276.7397 & 50.5 &  8121.5973 & 47.0  \\
5842.7295 & 22.5 &6303.5135 & 33.0 &            &       \\
\hline
  \end{tabular}
\end{table}

\subsection{Spectroscopy}

Between 2009 and 2018, HD\,25631 apparently served as a calibration source  (ESO Program ID 60.A-9022) for the X-Shooter spectrometer \citep{ver11} installed at Cerro Paranal Observatory on UT2 (until mid-2013) and UT3 (since then). This spectrometer provides medium-resolution data simultaneously in the near-UV, visible, and near-IR ranges. The main reduction steps (bias subtraction, flat-fielding, order extraction and merging) are done by a specific ESO pipeline and the reduced spectra can be downloaded from the public ESO archive science portal\footnote{http://archive.eso.org/scienceportal/home}. The star was observed at 71 epochs, often with multiple exposures over the same night. Generally, the exposures were consecutive, but in one case they were taken with a 5h interval. Since that interval still corresponds to a very small fraction of the stellar cycle (see below), all spectra taken on the same night were averaged to improve the signal-to-noise ratio (SNR), which then reached several hundreds. Note however that spectra showing wavelength gaps were discarded from this averaging process. Normalization was performed by fitting polynomials of low order to chosen continuum windows. After examining the spectra, we decided to focus on the 4200--5100\,\AA\ (with H$\gamma$, H$\beta$, and several He\,{\sc i} lines), He\,{\sc i}\,$\lambda$5876\,\AA, and the 6420--6750\,\AA\ (with H$\alpha$ and He\,{\sc i}\,$\lambda$6678\,\AA) regions as they contain the strongest lines. To get the radial velocities for each spectrum, a cross-correlation was made with a synthetic spectrum with $T_{\rm eff}=20$\,kK and $\log(g)=3.75$ from the TLUSTY model grid at solar metallicity \citep{lan07}, convolved to the observed projected rotational velocity. In that process, all wavelength ranges were attributed the same weight. The individual heliocentric corrections were of course added to the resulting velocities. The derived radial velocities, rounded to half a \kms, are provided in Table \ref{rv25631}. As those spectra were taken in similar conditions, the velocity errors, derived from the cross-correlation function width \citep{zuc03}, were also similar, about 0.5\,\kms. Note that only one signature, that of the fast-rotating B-star, is clearly detected in the spectrum: HD\,25631 is thus an SB1 system.

\begin{table}
  \scriptsize
  \caption{Radial velocities derived from optical (right) and UV (left) spectroscopy of HD\,191495. \label{rv191495}}
  \begin{tabular}{lc|lc}
    \hline
$HJD$ & $RV$& $HJD$ & $RV$ \\
$-2.45e6$& (\kms) & $-2.45e6$& (\kms)  \\
\hline                                               
1767.117 &--16.3$\pm$1.5 & 10044.911 & 13.0$\pm$1.5 \\
1768.159 &  10.2$\pm$1.5 & 10046.942 & --45.5$\pm$1.0\\
\hline
  \end{tabular}
\end{table}

Two spectra of HD\,191495 were acquired in April 2023 with the HEROS spectrograph on the 1.2\,m robotic TIGRE telescope in Mexico \citep{schm14}. The two spectra have SNRs of 50 and 100. The main reduction steps (bias subtraction, flat-fielding, order extraction and merging, combination of subexposures) are done by a specific pipeline. The subsequent normalization and cross-correlation were done as for HD\,25631, the TLUSTY template having $T_{\rm eff}=28$\,kK and $\log(g)=4.5$. The right part of Table \ref{rv191495} provides the derived velocities. 

\begin{table}
  \scriptsize
  \caption{Radial velocities derived for HD\,46485.  \label{rv46485}}
  \begin{tabular}{lcllcl}
    \hline
$HJD$ & $RV$& Instr/Ref & $HJD$ & $RV$& Instr/Ref\\
$-2.45e6$& (\kms) &  &$-2.45e6$& (\kms) &   \\
\hline                                               
3683.6399 &30.4 &ELODIE, M09  & 4810.1206 &30.0$\pm$1.0 &Espadons   \\
3740.6290 &32.0 &FEROS, M09   & 4811.1384 &39.5$\pm$1.0 &Espadons   \\
4406.6388 &25.8 &AURELIE, M09 & 5817.7335 &27.5$\pm$1.0 &FIES, B23  \\
4408.6487 &44.7 &AURELIE, M09 & 5971.7503 &41.0$\pm$1.0 &Espadons   \\
4409.6440 &47.9 &AURELIE, M09 & 6321.5570 &24.5$\pm$1.0 &FIES, B23  \\
4414.6506 &39.6 &AURELIE, M09 & 6322.5508 &29.0$\pm$1.0 &FIES, B23  \\
4417.6071 &41.8 &AURELIE, M09 & 8031.7073 &20.0$\pm$2.0 &STELLA, B23\\
4421.6261 &40.2 &AURELIE, M09 & 8035.6988 &37.0$\pm$2.0 &STELLA, B23\\
4472.4275 &38.8 &AURELIE, M09 & 8044.6718 &36.5$\pm$2.0 &STELLA, B23\\
4473.4997 &36.5 &AURELIE, M09 & 8047.6801 &43.5$\pm$2.0 &STELLA, B23\\
4474.4504 &17.7 &AURELIE, M09 & 8055.6956 &19.5$\pm$2.0 &STELLA, B23\\
4475.4754 &12.6 &AURELIE, M09 & 8532.5759 &40.5$\pm$1.0 &UVES       \\
4546.7298 &34.2 &ESPRESSO, M09& 9201.7160 &39.0$\pm$1.0 &HERMES, B23\\
4547.7396 &48.1 &ESPRESSO, M09& 9202.7632 &12.5$\pm$1.0 &HERMES, B23\\
4548.7335 &50.4 &ESPRESSO, M09& 9204.7375 &23.0$\pm$1.0 &HERMES, B23\\
4549.7386 &41.5 &ESPRESSO, M09& \\
4550.7509 &31.9 &ESPRESSO, M09& \\
\hline
  \end{tabular}
  
  {\scriptsize For the Instrument/Reference column, symbols are the following. M09: the Helium lines' velocities of \citet{mah09} were averaged and those from ESPRESSO further corrected for interstellar line velocity difference. B23: the velocities of spectra from \citet{bri23} were recalculated for several Helium lines using a cross-correlation technique, as for the newly presented Espadons and UVES spectra. 1-$\sigma$ errors, derived from the cross-correlation function width, are provided for all new measurements. See text for details.}
\end{table}

The apparent brightness of HD\,46485 is only 8.3\,mag in $V$, so the SNR of its spectra is much lower than for HD\,25631 ($V=6.5$\,mag). Nevertheless, spectra have been taken and used. The multiplicity study of \citet{mah09} relies on two public spectra, taken with the high-resolution spectrometers ELODIE (France, in November 2005, \citealt{bar96}) and FEROS (ESO/Chile, in January 2006, \citealt{kau99}) - both available from their respective archives, as well as on new spectra taken with AURELIE (France, in 2007--2008, \citealt{gil94}) and ESPRESSO (Mexico, in March 2008\footnote{http://www.astrosen.unam.mx/Ens/Instrumentacion/manuales/echelle/echelle.html}). These authors provided radial velocities for the He\,{\sc i}\,$\lambda$4471,5876\AA\ and He\,{\sc ii}\,$\lambda$4542,4686\AA\ lines, along with velocities of the interstellar Na\,{\sc i} lines. In Table \ref{rv46485}, averages of all Helium line velocities were made. Note that the Mexican data displayed lower values of the velocities for the interstellar lines, hence the Helium velocities were corrected for this difference. The abundance study of \citet{caz17obs} used the same ELODIE and FEROS spectra\footnote{Note that the velocity of the ELODIE spectrum is listed in the paper {\it without} inclusion of the heliocentric correction of 22.3\,\kms, making the actual velocity 33.0$\pm$7.3\,\kms, which agrees well with that of \citet{mah09}.}. Finally, the multiplicity study of \citet{bri23} relies on high-resolution spectra taken for IACOB \citep{sim15} with FIES in 2011--2013 \citep{tel14}, Stella in October 2017 \citep{web08}, and Hermes in December 2020 \citep{ras11}, all instruments installed at the Canary Islands Observatories. To increase precision, velocities were recalculated for those data by cross-correlating them with a synthetic spectrum with $T_{\rm eff}=30$\,kK and $\log(g)=3.5$ from the TLUSTY model grid \citep{lan03}. This was done on small intervals centered on the strongest Helium lines (He\,{\sc i}\,$\lambda$4471,5876\AA\ and He\,{\sc ii}\,$\lambda$4542\AA) as the larger noise for this star made cross-correlations on broad intervals more difficult and much less reliable. Results are provided in Table \ref{rv46485}. 

\begin{figure*}
  \begin{center}
    \includegraphics[width=8cm]{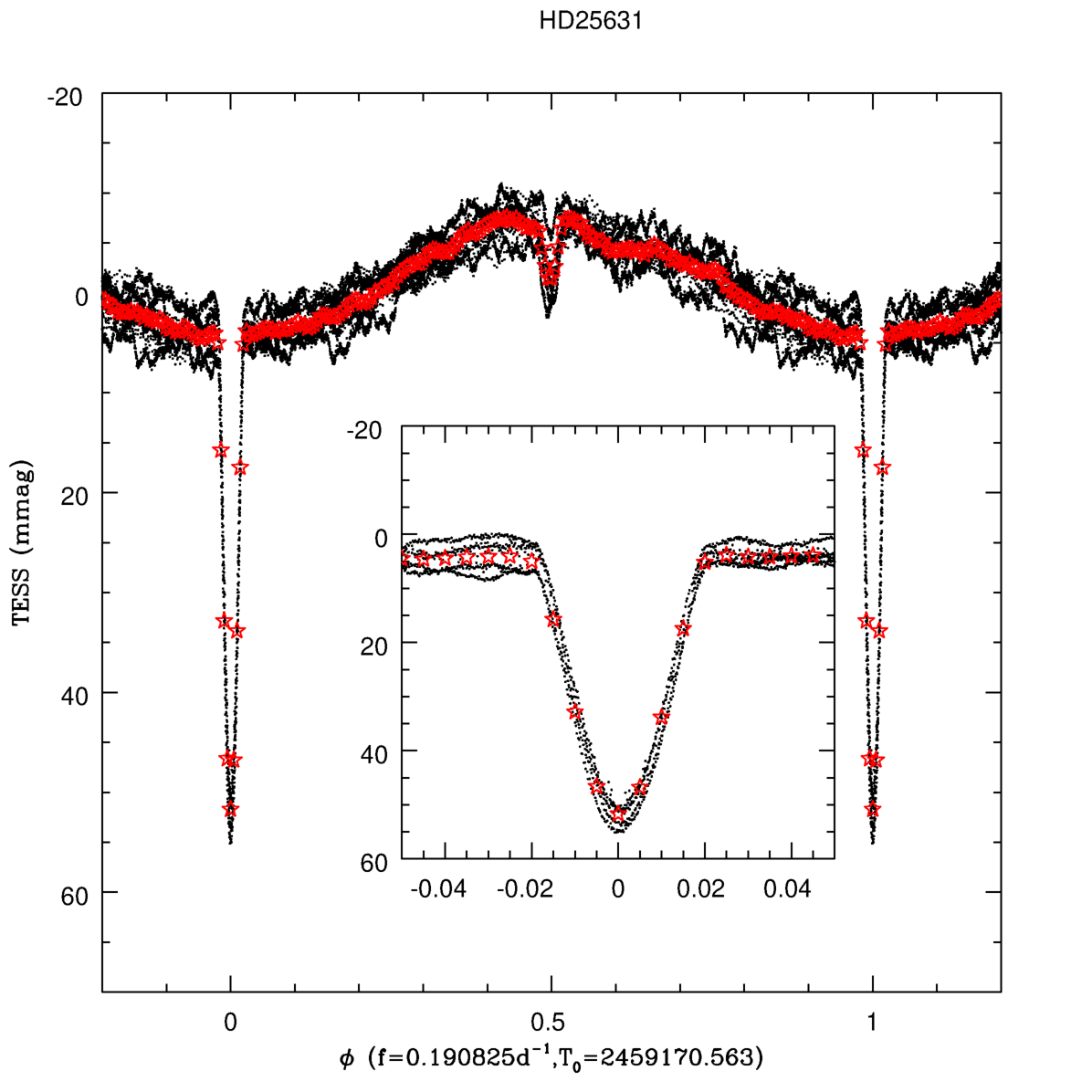}
    \includegraphics[width=8cm]{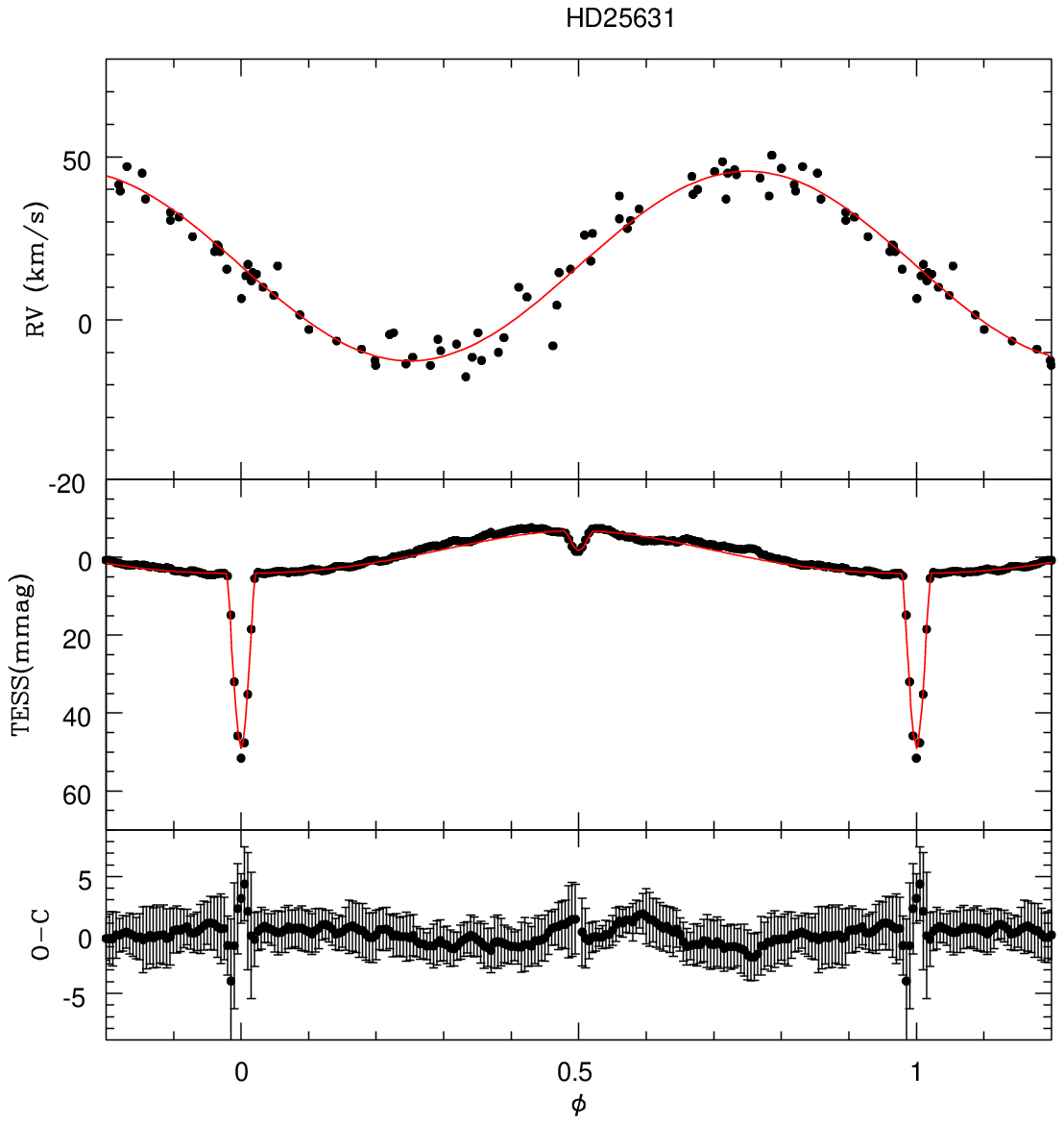}
  \end{center}
  \caption{{\it Left:} The \te\ light curves of HD\,25631, folded on the best-fit ephemeris. Data from Sectors 4, 5, and 31 are shown with black dots while the average curve combining all data is displayed with red stars. {\it Right:} Radial velocities of HD\,25631 with their best-fit orbital solution (top), average light curve, built using 200 bins from the \te\ data, compared to its best-fit model (middle), and their difference compared to the 1$\sigma$ dispersion in each bin (bottom). Because of the non-orbital variability of the source, these error bars exceed the nominal accuracy of the \te\ data.}
\label{25631rvphot}
\end{figure*}

\begin{figure*}
  \begin{center}
    \includegraphics[width=8cm]{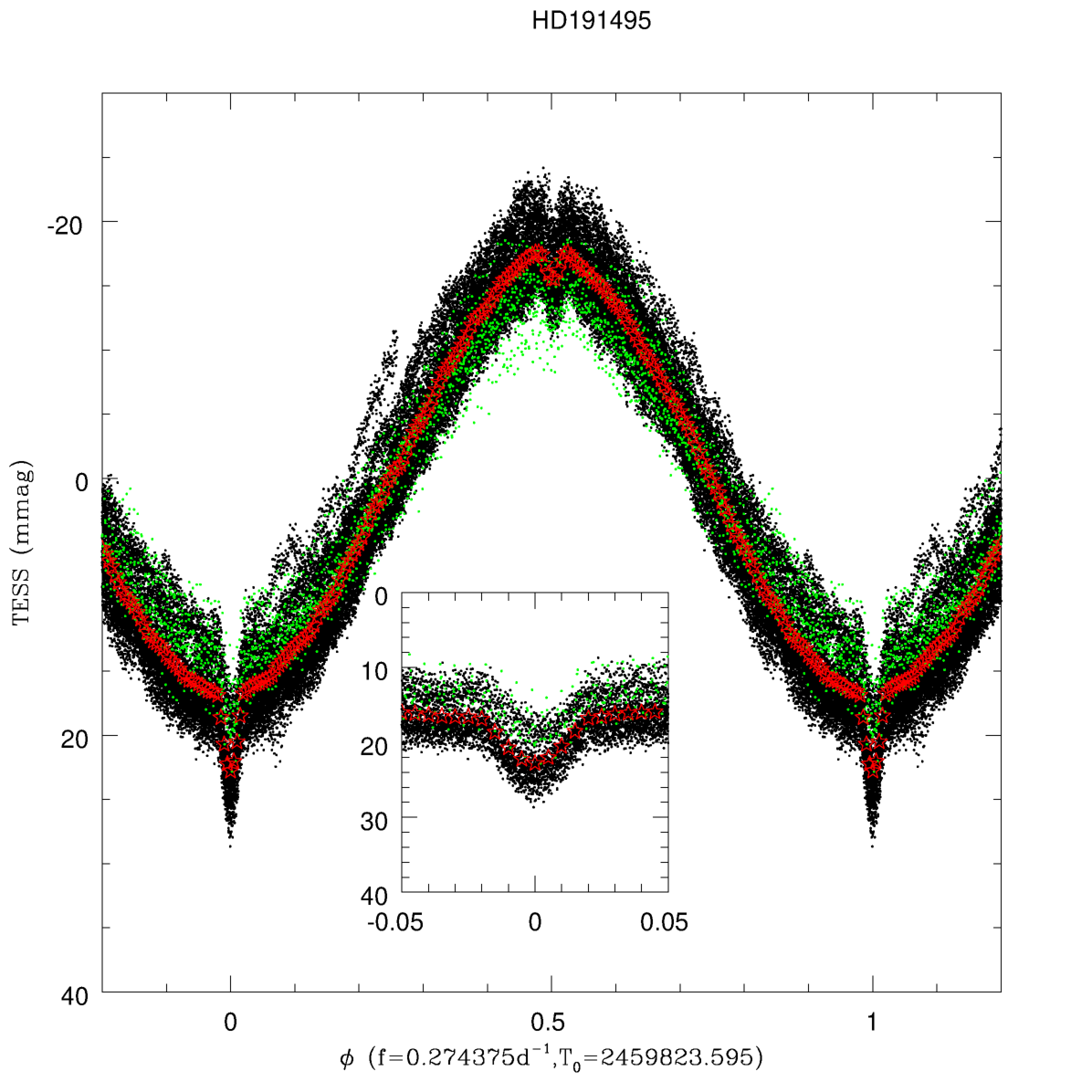}
    \includegraphics[width=8cm]{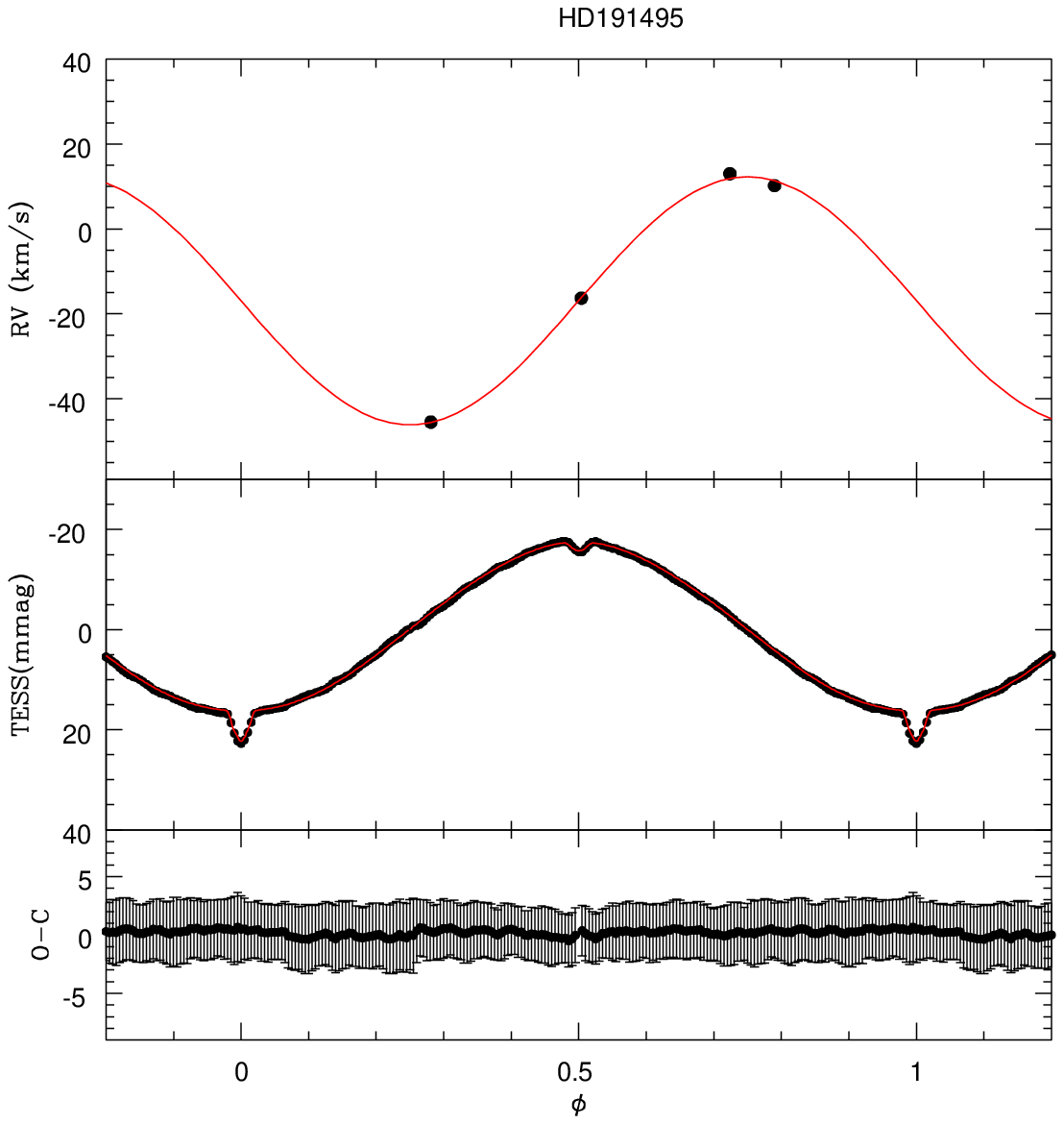}
  \end{center}
  \caption{Same as Fig. \ref{25631rvphot} for HD\,191495. The higher and lower cadence \te\ data are shown in the left panel with black and green dots, respectively.}
\label{191495fold}
\end{figure*}

\begin{figure*}
  \begin{center}
    \includegraphics[width=8cm]{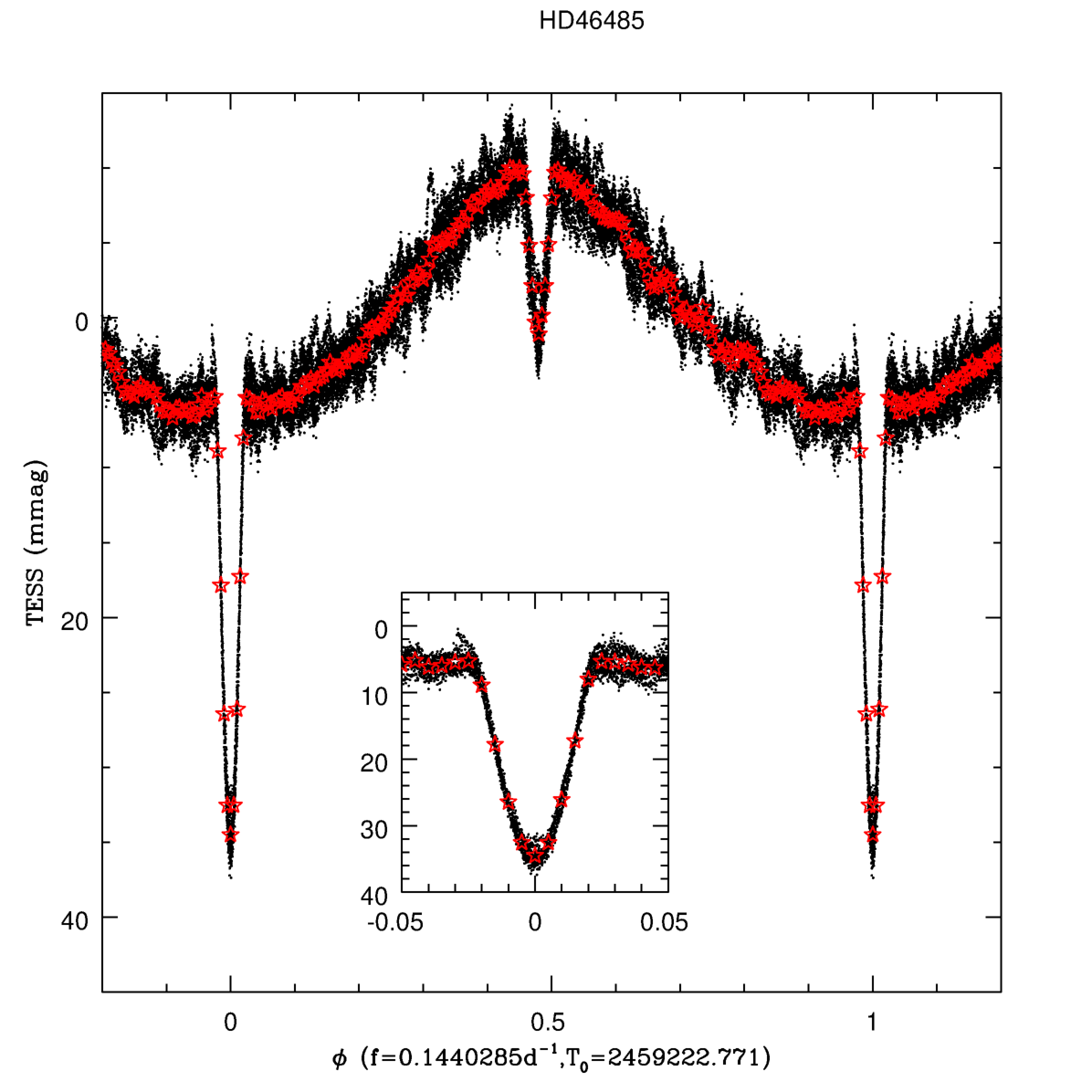}
    \includegraphics[width=8cm]{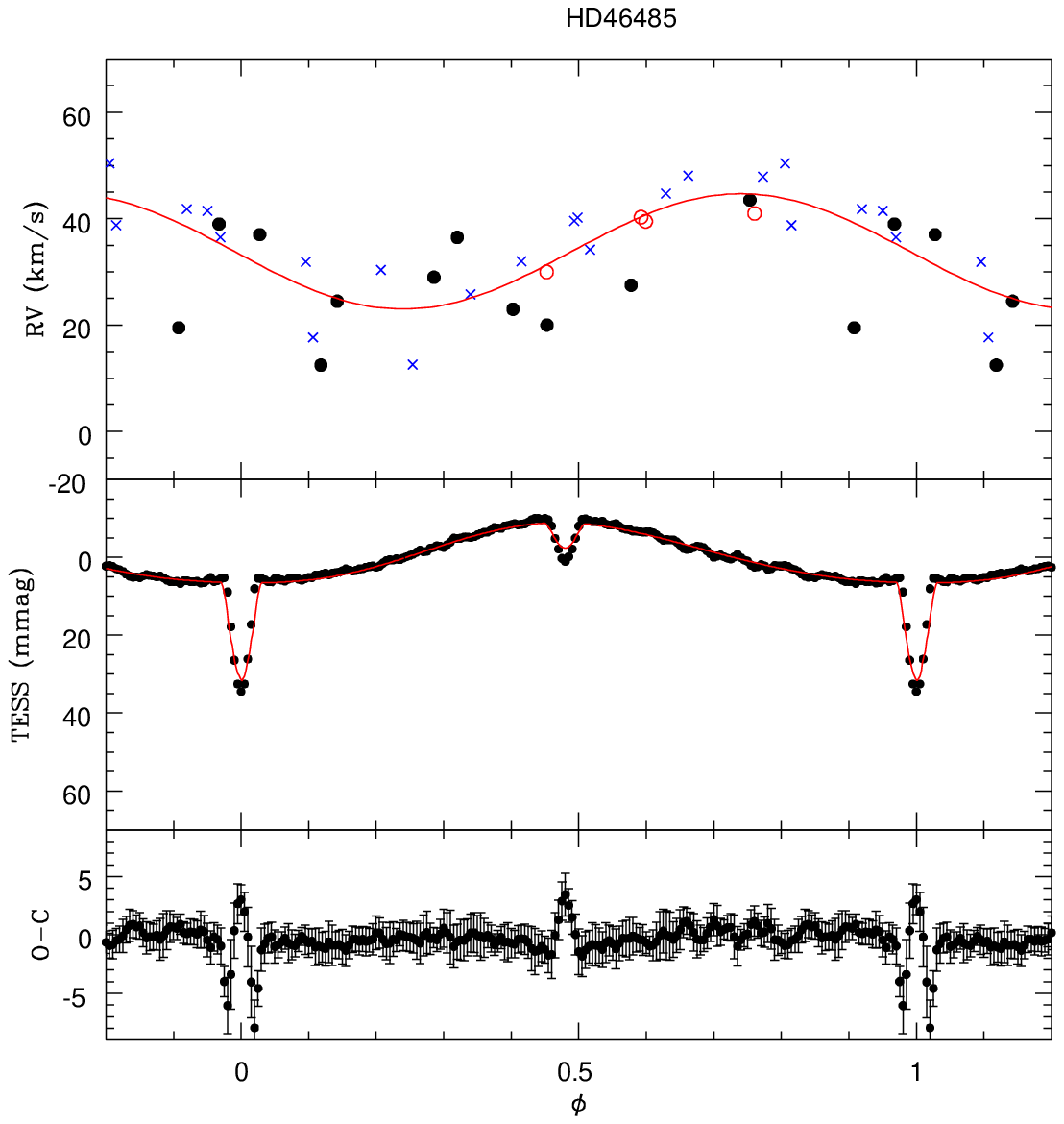}
  \end{center}
  \caption{{\it Left:} The \te\ light curves of HD\,46485, folded on the best-fit ephemeris. Data from Sectors 6 and 33 are shown with black dots while the average curve combining all data is displayed with red stars. {\it Right:} Radial velocities of HD\,46485 with their best-fit orbital solution (top - blue crosses, black dots and red circles correspond to velocities of M09, B23, and this work, respectively, see Table \ref{rv46485}); average light curve with its best-fit model (middle) and the O-C residuals compared to the 1$\sigma$ uncertainties (bottom). }
\label{46485rvphot}
\end{figure*}

In addition, high-resolution spectra of HD\,46485 were also taken at three epochs in 2008--2012 with the high-resolution spectrometer Espadons installed on the Canada-France-Hawaii Telescope \citep{man03} and at one epoch in February 2019 by the high-resolution spectrometer UVES installed at Cerro Paranal Observatory \citep{dek00}. The reduced spectra were downloaded from their public archives\footnote{http://polarbase.irap.omp.eu/ and http://archive.eso.org/scienceportal/home} and normalized as mentioned before. Note that, for each epoch, the Espadons data that we used correspond to averages of multiple consecutive subexposures taken with the aim of searching for magnetic fields. As before, radial velocities were derived by cross-correlating the observed spectra with an atmosphere model (see Table \ref{rv46485} for results). Note that there is also LAMOST data for this star \citep{li21}. The individual spectra can be downloaded from the archives\footnote{http://dr7.lamost.org/}. However, their low resolution prohibits deriving precise velocities as required by our study so they were not used. 

In the UV domain, a few spectra are available for HD\,25631 and HD\,191495. Four IUE \citep{bog78} low-dispersion spectra were taken for HD\,25631, two in the FUV (SNR$\sim$37 and $\sim$20 at 1500\,\AA) and two in the NUV (SNR$\sim$32 and $\sim$14 at 2500\,\AA). From the IUE handbook, the resolution is approximately $R\sim400$ in the FUV and $R\sim500$ in the NUV, compared to the high-dispersion mode which provides $R\sim10000$ in the FUV and $R\sim17000$ in the NUV. The two FUSE \citep{moo00} observations for HD\,191495 ($V=8.4$\,mag) were taken 1.04\,d apart, with exposure times of 2.56 and 2.52 hours, each with a SNR of $\sim$30 at 1150\,\AA; resolution is about 20\,000 for these spectra. Velocity shifts between these two spectra are obvious. We have fitted with Gaussians the two cleanest photospheric lines (the Si\,{\sc iv}\,$\lambda$1128.3\,\AA\ and C\,{\sc iii}\,$\lambda$1175.7\,\AA\ multiplets) and the left part of Table \ref{rv191495} provides the average of the derived velocities. For HD\,46485, the IUE data were already examined in \citet{wan18} hence we did not repeat their analysis. All these data were downloaded from the MAST archives.

\section{Results}
\subsection{Previous knowledge on the targets and evolutionary parameters}

\begin{table}
  \scriptsize
  \caption{Published parameters of the targets. \label{lit}}
  \begin{tabular}{lccc}
    \hline
    & HD\,25631 & HD\,191495 & HD\,46485 \\
    \hline
Sp. Type          & B3V [1,2,3] & B0V [5] & O7V((f))n var? [8] \\
$v\sin(i)$(\kms)  & 221 [2] & 201 [6] & 334$\pm$16 [9] \\
$T_{\rm eff,1}$(K)   & 19930 [2] & 28250$\pm$1300 [7] & 36100$\pm$700 [9] \\
$M_1$(M$_{\odot}$)   & 6.1$\pm$0.7 [3], 6.9$\pm$0.5 [4] & 12.3$\pm$1.1 [7] & \\
$R_1$(R$_{\odot}$)  & 4.1$\pm$0.8 [4] &  & 7.5$\pm$0.1 [9]\\
age (Myr)           & 23$\pm$11 [4] & \\
$\log(L_1/L_{\odot})$ & 3.33$\pm$0.19 [4] & 4.12$\pm$0.04 [7] & 4.93$\pm$0.03 [9] \\
\hline
  \end{tabular}
  
  {\scriptsize [1] \citet{pau05}, [2] \citet{bra12}, [3] \citet{hoh10}, [4] \citet{lyu02,lyu04} - luminosity for a distance of 402\,pc, [5] \citet{bow08,mel20}, [6] \citet{gle05}, [7] \citet{qui21}, [8] \citet{mai16}, [9] \citet{bri23} }
\end{table}

Table \ref{lit} summarizes the published parameters of our targets. HD\,25631 is a B3V field star with the latest spectral type of our sample. The most recent determination of its temperature and rotation rate was made by \citet{bra12}, which agree well with previous determinations \citep{lyu02,str05,pau05,hoh10}. Other properties were reported by \citet{hoh10,lyu02,lyu04}. Only \citet{jer77} reported on the photometric behaviour of this star, mentioning a maximum of 40\,mmag for the variations without identifying a periodicity. The B0V star HD\,191495 belongs to the NGC\,6871 cluster in Cyg\,OB3 (e.g. \citealt{rao20}). The age of this association was found to be around 11\,Myr but was thought to be lower, 3--5\,Myr for the early-type stars in the area \citep{dia21,qui21}. HD\,191495 was never intensively studied although some parameters were derived by \citet{qui21} using a simple SED fitting and its projected rotational velocity appears close to that of HD\,25631 \citep{gle05}. Furthermore, a sinusoidal variation of 28\,mmag peak-to-peak amplitude and period of 0.789\,d was reported for this star by \citet{zak95}. The authors proposed it to be linked to rotational modulation of what they assumed to be a magnetic Ap star. In fact, that period is a daily alias of the true period, whose value is clearly revealed by \te\ (see below). The O7V((f))n var? star HD\,46485 belongs to the young NGC\,2244 cluster (of age$\sim$2\,Myr, see \citealt{lim21} and references therein). It has been the target of multiple studies. The stellar parameters were usually derived from model atmosphere fitting, most recently by \citet{bri23} which agrees well with previous studies (e.g. \citealt{caz17obs}). Beyond its extremely fast rotation, no peculiarity was detected for this star: the surface abundances were found to be solar \citep{mar15,caz17obs,caz17theo} and the star was not found to be magnetic \citep{gru17,pet19} or peculiar in the X-ray range \citep{bro13}. Regarding multiplicity, no visual companion is reported for this star \citep{mas09,san14}, although a partially resolved companion is found in the analysis of HST-FGS data \citep{ald15}. From radial velocities only, HD\,46485 is often considered as presumably single \citep{mah09,caz17obs}. Its eclipsing nature was however discovered from the first \te\ dataset \citep{bur20} and reaffirmed in \citet{bri23}. In \citet{bur20}, the system, along with seven other eclipsing cases, was considered as a good candidate for the growing class of OB+compact companion systems. 

\begin{table}
  \scriptsize
  \caption{Properties of the targets. \label{results}}
  \begin{tabular}{lccc}
    \hline
    & HD\,25631 & HD\,191495 & HD\,46485 \\
    \hline
$\log(L_1/L_{\odot})$ & 3.24$\pm$0.02 & 4.51$\pm$0.04 & 4.95$\pm$0.03 \\
$d$(pc)             & 376$\pm$9 & 1607$\pm$70 & 1248$\pm$40 \\
\hline
\multicolumn{4}{l}{\it Evolutionary models (BONNSAI)}\\
$M_1^{evol}$(M$_{\odot}$)   & 6.8$\pm$0.3 & 15.2$\pm$0.8 &  24.2$\pm$1.2\\
$R_1^{evol}$(R$_{\odot}$)  & 3.6$\pm$0.3 & 7.2$\pm$0.9 & 7.2$\pm$0.6 \\
age$^{evol}$ (Myr) & 9.6$\pm$9.4 & 7.6$\pm$1.8 & 2.5$\pm$1.1  \\
\hline
\multicolumn{4}{l}{\it Spectroscopic solutions}\\
$e$ & 0 & 0 & 0$^*$ \\
$\gamma_1$ (\kms) & 16.5$\pm$0.6 & --17.5$\pm$1.0 & 33.9$\pm$1.2 \\
$K_1$ (\kms) & 29.1$\pm$0.8 & 29.4$\pm$1.0 & 10.8$\pm$1.7 \\
$f(m)$(M$_{\odot}$) & 0.0133 & 0.0094 & 0.00091 \\
                   & $\pm$0.0012 & $\pm$0.0010 & $\pm$0.00044 \\
\hline
\multicolumn{4}{l}{\it Light curve models}\\
$M_1$(M$_{\odot}$),adopted   & 7--8 & 15 & 24 \\
$q$,adopted        & 0.12--0.15 & 0.08--0.12 & 0.035--0.045 \\
$T_0$ (deepest ecl.)$^\dagger$ & 2459170.563 & 2459823.595 & 2459222.771 \\
$1/P$ (d$^{-1}$)$^\dagger$    & 0.190825 & 0.274375 & 0.1440285 \\
$e$ & 0 & 0 & 0.033 \\
$i$($^{\circ}$)     & 79--80 & 67--69 & 73--76 \\
$R_1$(R$_{\odot}$)  & 4.0--4.4 & 6.5--7.1 & 10.--12.\\
$f_1$             & 0.29--0.34 & 0.48--0.53 & 0.39-0.47 \\
$M_2$(M$_{\odot}$)  & 0.8--1.2 & 1.2--1.8 & 0.8--1.1 \\
$R_2$(R$_{\odot}$)  & 1.6--1.9 & 3.0--3.2 & 2.8--4.1\\
$f_2$             & 0.30--0.35 & 0.61--0.71 & 0.48--0.69 \\
$T_{\rm eff,2}$(K)   & 3400--6000 & 5600--11000 & 5900--15000 \\
\hline
  \end{tabular}
  
  {\scriptsize $f_i$ are Roche Lobe filling factors. $^*$ indicates that, while eccentricity actually is small but non-zero, similar results are achieved for the spectroscopic solution if considering a circular orbit - see text for details. $^\dagger$ indicates that the errors on these values is about 1 on the last decimal digit, except for the frequency of HD\,46485 which is 0.1440285$\pm$0.000005\,d$^{-1}$.}
\end{table}

We have determined the bolometric luminosities of the targets using their Gaia distances \citep{bai21}, a bolometric correction adapted to the chosen temperature \citep{ped20}, their known magnitudes ($V=6.45, 8.42, 8.27$ for HD\,25631, HD\,191495, and HD\,46485, respectively) and reddenings (negligible for HD\,25631, $A_V=1.12$\,mag for HD\,191495 - \citealt{qui21}, $E(B-V)$=0.65\,mag for HD\,46485). The derived values are listed in Table \ref{results}. For HD\,25631, this luminosity agrees with previous determinations \citep{lyu02} and with expectations for stars of that temperature, although it is at the highest end of expectations for this spectral type. The luminosity of HD\,191495 also agrees well with expectations for B0V stars, while that of HD\,46485 is only slightly below that expected for an O7V star.

Using the known temperatures and new luminosities, we searched for the best-fit evolutionary track using BONNSAI\footnote{https://www.astro.uni-bonn.de/stars/bonnsai/} \citep{schn14} with parameters adapted to the Milky Way \citep{bro11}. The derived values are listed in Table \ref{results} with the ``evol'' superscript. 

\subsection{Spectrocopic analysis}
\subsubsection{Orbital solutions}
The radial velocities were derived in Section 2.2. Velocities of only one component could be determined so our targets are thus SB1 systems. Using the photometric periods (see next section), we establish their SB1 orbital solution thanks to the Li\`ege Orbital Solution Package \citep[LOSP,][]{San06}. We tested both circular and eccentric orbital solutions, and the best-fit parameters are listed in Table \ref{results}.

For HD\,25631, an eccentric solution did not improve the quality of the fits (rms of 4.9\,\kms) and the best-fit eccentricity $e = 0.048 \pm 0.029$ was small and only marginally (1.5$\sigma$) significant. In addition, the light curve, with the eclipses centered on $\phi=$0 and 0.5 does not reveal any sign of non-zero eccentricity (see below) hence a circular orbit was thus finally assumed (Fig. \ref{25631rvphot}). \citet{vie17} and \citet{hoh10} quote typical masses of B3V stars as 7.7 and 6.6\,M$_{\odot}$, respectively, in line with BONNSAI results (Table \ref{results}). We therefore adopted a probable mass value of 7--8\,M$_{\odot}$. The mass function then indicates that the mass ratio $q=M_2/M_1$ should be at least 0.128--0.135 for $M_1$=7--8\,M$_{\odot}$. Accounting for uncertainty on the B-star mass and $i>60^{\circ}$, $q$ most probably lies between 0.12 and 0.15 and this range was used for the photometric analysis (see next section).

For HD\,191495, the FUSE data revealed a velocity shift. Those data were taken 8000\,d before the \te\ data but, thanks to the high precision on the photometric frequency (see below), a one-sigma change would lead to a change in phase below 0.01. The more recent TIGRE data appear fully in line with them (Table \ref{rv191495} and Fig. \ref{191495fold}). We have fitted a simple sinusoid to the derived velocities assuming the above ephemeris (Fig. \ref{191495fold}). For the mass of the star, a value of $\sim$15\,M$_{\odot}$ appears typical for its spectral type (e.g. \citealt{hoh10}, \citealt{vie17}) and agrees with BONNSAI fitting (Table \ref{results}). The derived mass function then leads to a minimum mass ratio $q=0.09$. In view of the limited amount of spectroscopic data, we decided to explore a reasonably broad range of mass ratios between 0.08 and 0.12 for the photometric fitting.

HD\,46485 is the only case displaying a clear eccentricity as revealed by the phase shift between the two eclipses (see next section). However, the high scatter in the RVs does not enable us to constrain the (small) eccentricity well from the velocity curve (Fig. \ref{46485rvphot}). In fact, the best-fit circular and eccentric solutions yield similar main parameters, although with larger errors in the latter case\footnote{Allowing for a non-zero eccentricity yields $e = 0.02^{+0.20}_{-0.02}$ and $f(m) = 0.00090 \pm 0.00054$\,M$_{\odot}$.}. The mass function implies that mass ratios of 0.035--0.040 are expected for $M_1=24$\,M$_{\odot}$ and $i>60^{\circ}$. Allowing for uncertainties, we increased the range to 0.035--0.045 for the light curve modelling. Test calculations with {\sc Nightfall} indicate that the phase shift of the eclipses in the light curve requires a minimum eccentricity of $0.033$, which is fully consistent with the radial velocity solution.

\subsubsection{Search for hot companions}

To interpret correctly the light curves, it is crucial to know whether the fast-rotating OB stars are the hottest or coolest objects of the systems. In this context, we searched for the signatures of a hot stripped star of sdO type. To this aim, a cross correlation function (CCF) analysis was performed similar to the methods of \citet{wan18}. Synthetic spectra were generated using the non-LTE radiative transfer code CMFGEN \citep{hil98} for $T_{\rm eff}$=40, 45, and 50\,kK and $\log(g)=4.5$. These spectra were then broadened to correspond to $v\sin(i)$ values between 1 and 120\,\kms. The observed spectra were smoothed with what is essentially a high-pass filter, by convolving the data with a Gaussian kernel with a width (standard deviation) equal to 2 pixels in order to remove wide features (i.e. those from the rapid rotator). The result was used to ``normalize'' the spectra and the synthetic spectra were then cross-correlated to these ``pseudo continuum normalized'' data. Note that this method only works if the projected rotational velocity of the hot companion is significantly less than that of the primary star, which is the case in virtually all observed Be+sdO systems and also consistent with theoretical predictions. In \citet{wan21}, detected sdO stars display upper limits on the projected rotational velocity of 15--36\,\kms, the single outlier being the hot companion of HD\,51354 with $v\sin(i) = 102$\,\kms\ which is still 3$\times$ lower than that of its Be star primary.

For HD\,25631, the cross-correlation function (CCF) was built for the highest-SNR spectrum of each range considering the wavelength intervals 1241--1950\,\AA\ (FUV) and 2000--3100\,\AA\ (NUV). The CCF analysis did not result in any detection of a hot slower rotator, with a limit on flux ratio $F_2/F_1$ of about 10\%. For HD\,191495, no sharp photospheric features from a putative hot sub-luminous more slowly rotating companion (i.e. an sdO star) are evident by-eye in the FUSE observations. Prior to cross-correlating the synthetic spectra to the FUSE data, interstellar absorption lines and bands were clipped out. The synthetic spectra were then cross-correlated to the ``pseudo continuum normalized'' data in the range between 1105--1183\,\AA\ (blueward of this, the spectrum is dominated by interstellar absorption). The CCF analysis did not reveal any hint of an sdO star in the spectra. Smaller wavelength ranges were analyzed in the same way, providing the same null results. The upper limit on the flux ratio $F_2/F_1$ is here below 5\% for typical slow-rotating hot companions (or 10--20\% if considering atypical companions with projected rotational velocities of 50--100\,\kms). For HD\,46485, the absence of detection of a hot, stripped-star signature from UV data was already reported by \citet{wan18}. This strongly suggests that, in all cases, the companion is most probably cooler than the fast-rotating massive star. 

\subsubsection{Uncovering the secondary spectrum}
Among our three targets, only one has a sufficient number of high-quality spectra to attempt for reconstructing and uncovering the spectrum of the secondary star: HD\,25631. To this aim, we restricted ourselves to those 49 X-Shooter spectra that do not suffer from any obvious instrumental problems (such as fringes or problems at echelle order connection). We used our spectral disentangling code based on the shift-and-add method described by \citet{GL}. To maximize our chances to detect the heavily diluted signature of the companion, we applied this technique to the near-IR spectral domain between 8400 and 8800\,\AA. Not only does it contain the Paschen lines which are ubiquitous in many stars but, in low-mass main-sequence stars, this spectral range is also dominated by the strong absorptions of the near-IR Ca\,{\sc ii} triplet at 8498, 8542 and 8662\,\AA\ \citep[e.g.][]{Mun99}.   

\begin{figure}
\begin{center}
    \includegraphics[width=8.5cm]{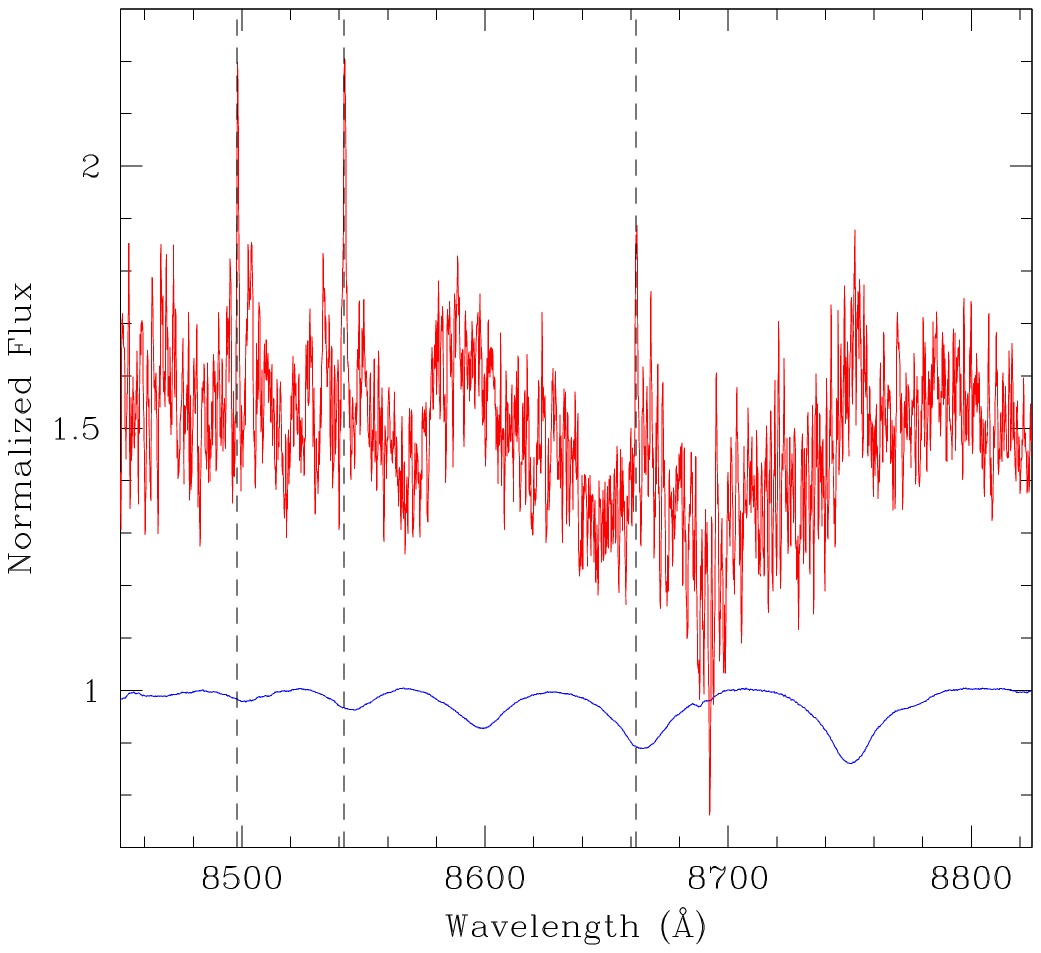}
\end{center}  
\caption{Disentangled normalized spectra of the primary (blue, bottom) and secondary (red, top) star of HD\,25631, assuming a mass ratio of $q = 0.135$, and a brightness ratio of 0.0067. The reconstructed secondary spectrum was corrected for the dilution by the light of the primary and was shifted vertically upwards by 0.5. The dashed lines indicate the wavelengths of the near-IR Ca\,{\sc ii} triplet.}
\label{disent}
\end{figure}

The radial velocities of the primary star were set to the best-fit circular orbital solution
$$RV_p (\phi) = 16.5 - 29.1\,\sin{(2\,\pi\,\phi)}$$  
whilst those of the secondary star were computed as
$$RV_s (\phi) = 16.5 + \frac{29.1}{q}\,\sin{(2\,\pi\,\phi)}$$
We explored values of $q$ between 0.12 and 0.15, thus sampling the most likely range of values of this parameter. We performed a total of 30 iterations to disentangle the spectra, although convergence of the spectral shape was usually reached well before. According to our light curve solutions (see below), it should be noted that the brightness ratio between the secondary and the primary over this waveband reaches values between 0.003 and 0.01. The reconstructed and renormalized (assuming a brightness ratio of 0.0067) spectra of the primary and secondary star for $q=0.135$ are shown in Fig.\,\ref{disent}. Whilst the primary spectrum is dominated by the strong H\,{\sc i} Paschen lines, as expected, the secondary spectrum unveals no absorption lines at the wavelengths of the Ca\,{\sc ii} triplet. Instead, we find narrow emission lines at these wavelengths. This does not depend on the chosen value for $q$, but holds true for all values that we have considered.

To assess our ability to uncover the secondary's spectral features, we applied the same disentangling technique to a sample of synthetic spectra generated by combining synthetic spectra from \citet{Mun00} and \citet{Cas01} for effective temperatures of 5750\,K for the secondary and 19500\,K for the primary. We assumed brightness ratios between 0.003 and 0.01 and a SNR ratio of individual spectra of 150, in line with that of our data in that range. The primary and secondary spectra were shifted according to the best-fit orbital solution considering $q = 0.135$. The sampling of the orbital cycle was set to be identical to that of the actual observations. The result was that, although the SNR of the reconstructed secondary spectra was low (between 3 and 10), the Ca\,{\sc ii} absorptions were clearly uncovered, even for a brightness ratio as low as 0.003. Hence, the SNR of the actual observations cannot explain the lack of Ca\,{\sc ii} absorptions in the reconstructed secondary spectrum.  

At this stage, we note that there is an important caveat in the application of spectral disentangling to a system such as HD\,25631. Indeed, due to the illumination by the B-type star, the side of the secondary that faces the B-star is significantly hotter than the opposite hemisphere (this is the origin of the reflection effect detected in photometry). In the spectral disentangling, the reconstructed spectra represent a mean of the contributions from both hemispheres, which could smear out some spectral features. However, as becomes clear from Fig.\,2 of \citet{Mun00}, the Ca\,{\sc ii} absorptions dominate the spectrum of cool main-sequence stars over a very wide range of temperatures from 3500 to 7500\,K.

We have also attempted spectral disentangling around the H$\gamma$, H$\beta$ and H$\alpha$ Balmer lines. However, because of a less favorable brightness ratio, the results were much less conclusive. We find weak absorption lines at the wavelength of the Balmer lines, indicating that the secondary star is indeed a cool object. Yet, the SNR of the reconstructed spectra does not allow us to perform a spectral classification. We stress here that, by definition, the disentangled spectra represent a mean of the contributions of the two hemispheres of the secondary star, which would make it difficult to assign a well-defined spectral type. Finally, the Li\,{\sc i} $\lambda$\,6708 line is also most probably present. 

\subsection{Photometric analysis}

The \te\ light curves of the three stars appear strikingly similar: a regular sinusoidal variation, superimposed on narrow eclipses (Fig. \ref{lcall}). Such light curves indicate binarity beyond doubt. To pinpoint the periodicity, period search techniques \citep{hmm} were applied on individual as well as on combined light curves, and results agree. We then folded the light curves with the derived ephemeris, but eclipses appeared not exactly in phase, which allowed us to further refine the values: Table \ref{results} provides the best orbital frequencies and best times $T_0$ of deeper eclipse (i.e. with B-star behind), while Figures \ref{25631rvphot}, \ref{191495fold}, and \ref{46485rvphot} show the folded \te\ photometry. 

Examining the light curve shapes, the shallow sinusoidal variations could at first be considered as ellipsoidal variations. In such a case, there would be two cycles per orbit: the maximum brightness should be detected at quadratures and the minimum one at conjunctions. However, the eclipses necessarily correspond to conjunctions and they appear in all three cases at both extrema of the sinusoidal component of the light curves. This implies that the sinusoidal variations are not ellipsoidal in nature but are rather linked to reflection effects. In fact, the cooler star in the system has its hemisphere facing its hotter companion being brighter due to illumination and heating by the strong light emission of the latter star. This is corroborated by the fact that the deepest eclipse, i.e. that corresponding to the eclipse of the hotter object of the system, occurs at a brightness minimum, i.e. when the cooler object is in front with its brighter hemisphere hidden from view. 

A simple global fitting for such light curves has been proposed by \citet{moe15} - see their Eq. (3). It consists in fitting a sinusoidal variation to reproduce the reflection effect plus two Gaussians for the eclipses. The results of such a fitting for our stars is provided in Table \ref{moefit}. Note that the phase of the shallower eclipse, with respect to the deepest one, is compatible with 0.5 for both HD\,25631 and HD\,191495, suggesting a null eccentricity, while the 0.02 offset found for HD\,46485 indicates an eccentricity of at least $e_{min}=0.02\pi/2=0.03$.

\begin{table}
  \scriptsize
  \caption{Simple fitting of the light curves. \label{moefit}}
  \begin{tabular}{lccc}
    \hline
Name & Ampl. (mmag) & Width ($10^{-3}$) & Phase\\
\hline                                               
HD\,25631 & 49.4$\pm$0.6& 9.02$\pm$0.12\\
          &  5.8$\pm$0.6& 8.92$\pm$1.01&0.4976$\pm$0.0010\\
          & 11.3$\pm$0.2\\
HD\,191495&  6.2$\pm$0.2& 9.29$\pm$0.42\\
          &  1.6$\pm$0.3& 5.65$\pm$1.26&0.5017$\pm$0.0012\\
          & 33.7$\pm$0.1\\
HD\,46485 & 29.2$\pm$0.8& 9.80$\pm$0.32\\
          &  9.8$\pm$0.8& 9.29$\pm$0.91&0.4800$\pm$0.0009\\
          & 15.5$\pm$0.3\\
\hline
  \end{tabular}
  
  {\scriptsize For each star, the first line provides the amplitude, in mmag, and the Gaussian width, in thousandths of phase, of the deepest primary eclipse ($\Delta I_1$ and $\Theta_1$ in Eq. 3 of \citealt{moe15} - this eclipse occurs at $\phi=0$ by definition); the second line provides the amplitude, in mmag, the Gaussian width, in thousandths of phase, and the phase of the shallowest secondary eclipse ($\Delta I_2$, $\Theta_2$, and $\Phi_2$ in Eq. 3 of \citealt{moe15}); the third line provides the peak-to-valley amplitude of the sinusoidal reflection effect ($\Delta I_{refl}$ in Eq. 3 of \citealt{moe15}).}
\end{table}

The question then is: are the fast-rotating massive stars the cool or the hot components in these systems? The deeper eclipse corresponds to the hotter star being eclipsed, i.e. with the cooler star in front of the system as seen from Earth. If that cool star was the fast rotator, one would then see its radial velocity increasing afterwards as its orbital motion then takes it away from us. In a similar way, the radial velocity of the fast rotator would decrease after the shallower eclipse as it would then approach us. Figures \ref{25631rvphot}, \ref{191495fold}, and \ref{46485rvphot} show both photometry and radial velocities folded with the same ephemeris. As is obvious from those plots, the fast rotator velocity decreases around the deeper eclipse and increases around the shallower one. The fast rotator star therefore is the hotter component of the system, in agreement with the null result found from UV data. 

Everything is now ready for the photometric fitting. As the massive stars display additional photometric variations apparently not commensurate with the orbital period (see below), we averaged the light curves over phase, using 200 phase bins, and calculated the dispersion around this average. For HD\,25631, this average was done considering all data or considering only the highest-cadence data, which are corrected for crowding and other effects. Both methods yield similar results and averaging more cycles leads to a larger damping of the intrinsic stellar noise. For HD\,191495, this average was done using only the highest cadence data, as some slight magnitude difference during eclipses is detected for the low-cadence data (which are uncorrected, e.g. for contamination).

We analysed these average light curves with the {\sc nightfall} code (version 1.92) developed by R.\ Wichmann, M.\ Kuster and P.\ Risse\footnote{This code is available at http://www.hs.uni-hamburg.de/DE/Ins/Per/Wichmann/Nightfall.html} \citep{Wichmann}. Band $I$ was assumed for photometric data, as it shares a similar center as \te\ bandpass. This software uses the Roche potential to describe the shape of the stars. For a binary system with circular orbit and without any surface spots, the model is described by six parameters: the mass-ratio $q$, the orbital inclination $i$, the primary and secondary filling factors ($f_1$ and $f_2$ defined as the ratio between the stellar polar radius and the polar radius of the associated Roche lobe), and the primary and secondary effective temperatures ($T_{\rm eff,1}$ and $T_{\rm eff,2}$). For eccentric binaries, two additional parameters are required: the orbital eccentricity $e$ and the argument of periastron $\omega$. For such systems, the filling factors are defined with respect to the Roche lobe at periastron and the sizes of the Roche lobes are assumed to scale instantaneously with the actual orbital separation. For each system, we fixed the primary star's temperature (Table \ref{lit}) and its mass as well as the orbital period (Table \ref{results}), then we tested a grid of (fixed) values of $q$ (see previous section). Therefore, we were left with four free parameters in the case of circular orbits. We adopted a quadratic limb-darkening law, and reflection effects were accounted for by considering the mutual irradiation of all pairs of surface elements of the two stars \citep{Hendry}.

\begin{figure}
\begin{center}
    \includegraphics[width=8.5cm,bb=18 145 550 380, clip]{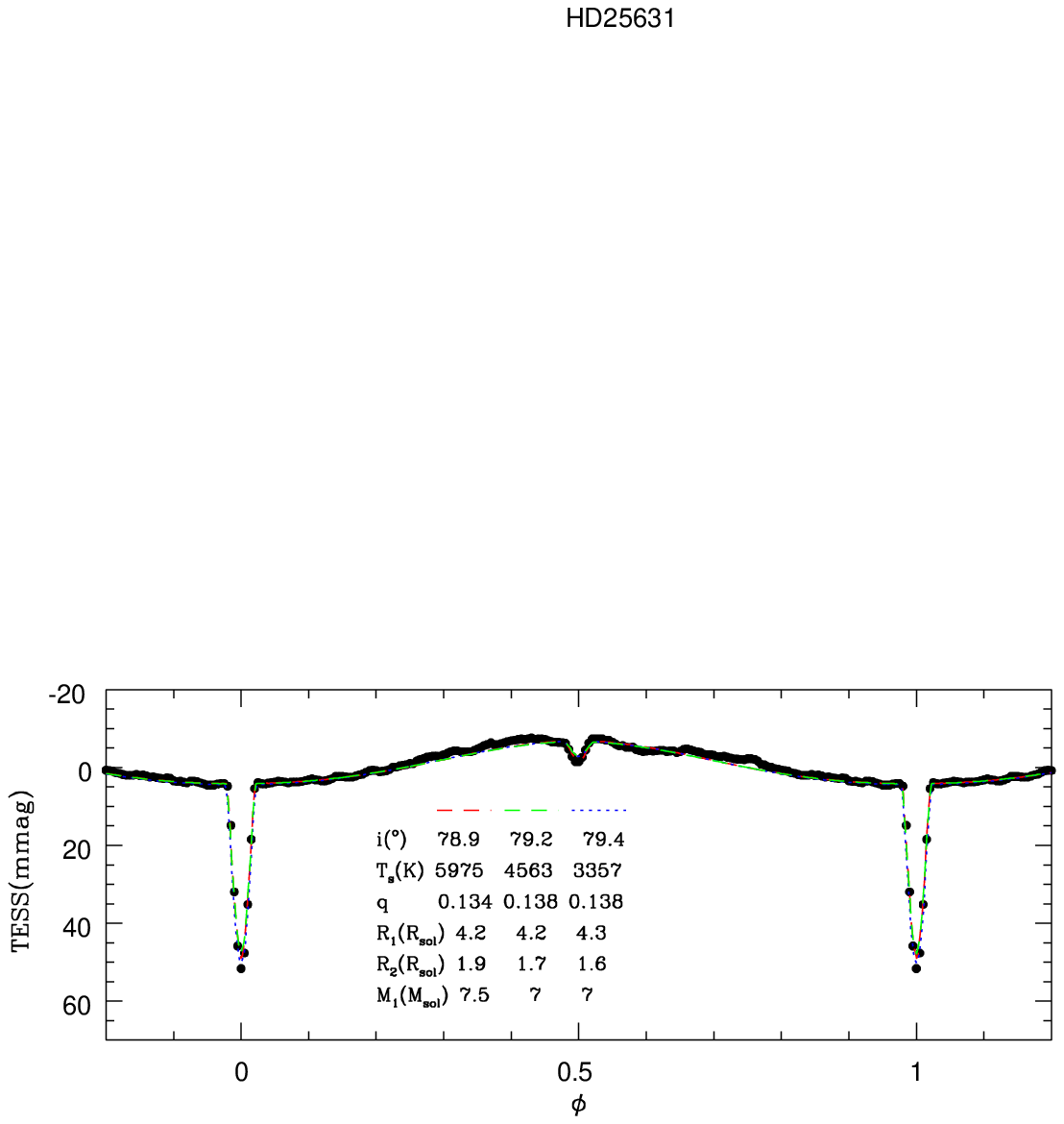}
\end{center}  
\caption{Comparison of models with the average \te\ light curve of HD\,25631, demonstrating that various combinations of model parameters yield theoretical light curves in agreement with the data.}
\label{ex}
\end{figure}

In all three cases, good fits were finally achieved (Figures \ref{25631rvphot}, \ref{191495fold}, and \ref{46485rvphot}). The fitting process is mainly driven by the sinusoidal variation, which affects all phases (hence concerns by far the largest number of points) and is excellently reproduced in all cases, but the narrow eclipses also appear reasonably fitted. Indeed, the $O-C$ residuals of our fits are usually consistent with zero within the errors (see bottom panels of Figures \ref{25631rvphot}, \ref{191495fold}, and \ref{46485rvphot}). This situation then leads to very small $\chi^2_{\nu}$ values of 0.325, 0.017 and 0.546 for the fits displayed in the figures. As one can clearly see for HD\,25631 and HD\,46485, the non-orbital variations clearly induce residual variations about the orbital modulations. The most significant deviations, although at $<3\sigma$, occur for a few data points during the eclipses of HD\,46485. In this context, before discussing the results, the modelling assumptions and limitations may be recalled. To first order, the amplitude of the sinusoidal modulation is mostly ruled by the filling factor of the secondary and its albedo coefficient. In the {\sc nightfall} code, the bolometric albedo coefficient is taken by default to be 0.5 for stars with a convective envelope ($T_{eff} < 7700$\,K) and 1.0 for stars with a radiative envelope. Hence, outside this transition temperature range, reflection is not expected to depend strongly on the secondary temperature. On the contrary, the relative depth of the eclipses is expected to depend on the ratio between the secondary and primary temperatures. However, since the systems analysed in the present work display partial or even grazing eclipses, the details of the limb-darkening law also play a key role in the eclipse modelling. The strong heating that the secondary's hemisphere facing the primary undergoes due to the irradiation by the primary has certainly a non-negligible impact on the secondary's atmosphere. This situation could alter the secondary's limb-darkening law, hence impacting the fitting of the secondary eclipses. The heating could also impact the albedo coefficient, and therefore in turn also the reflection effect as well as the secondary filling factor. Such effects are not considered in the existing light curve fitting tools and would require detailed hydrodynamic simulations of stars with strong illumination on one side. A full analysis of these effects is beyond the scope of our present paper, but they can most probably account for most of the small imperfections of our light curve fits. 

Exploring the parameter space led us to discover that several solutions actually provide fits of similar quality (see an example in Fig. \ref{ex}). Despite this, the best-fit inclination and stellar radii appear quite robust, with only small changes between various solutions. In contrast, the secondary temperature (hence its luminosity) appears somewhat ill-defined. To allow for this uncertainty which exceeds formal errors of any given individual solution, Table \ref{results} provides the range of values covered by the various valid solutions.

Knowing orbital periods and stellar radii, one may calculate the rotation rate associated to synchronization of rotation and orbital motion. The derived values are $v_{rot,sync}\sim40$, 94, and 80\,\kms\ for the primary stars of HD\,25631, HD\,191495, and HD\,46485, respectively. In all cases, the observed projected rotational velocities (Table \ref{lit}) are much larger than these values: the rotation is certainly not synchronized on the orbital motion and the massive stars truly rotate fast.

For HD\,25631, the lowest $\chi^2$ is achieved for $q=0.135-0.145$, which is in line with the mass ratio derived from the mass function using the best-fit inclination ($q=0.1375$). For HD\,191495, the best fit is formally found for $q=0.10$, which coincides well with that derived from the mass function for the best-fit inclination. For HD\,46485, we tried two values of the orbital eccentricity: 0.033 (the minimum eccentricity) and twice this value. Adopting the larger eccentricity value results in a reasonable fitting of the overall modulation due to the reflection effect but fails to reproduce well the eclipse of the secondary star. This suggests that the lower value of the eccentricity should be favored. Indeed, allowing even higher values of $e$ (e.g.\ the upper limit found from the RV solution) leads to the appearance of a so-called heartbeat signal in the light curve around periastron passage. Such a signal is clearly not observed in the data, thus ruling out values of $e > 0.1$.

\begin{figure}
  \begin{center}
    \includegraphics[width=8cm]{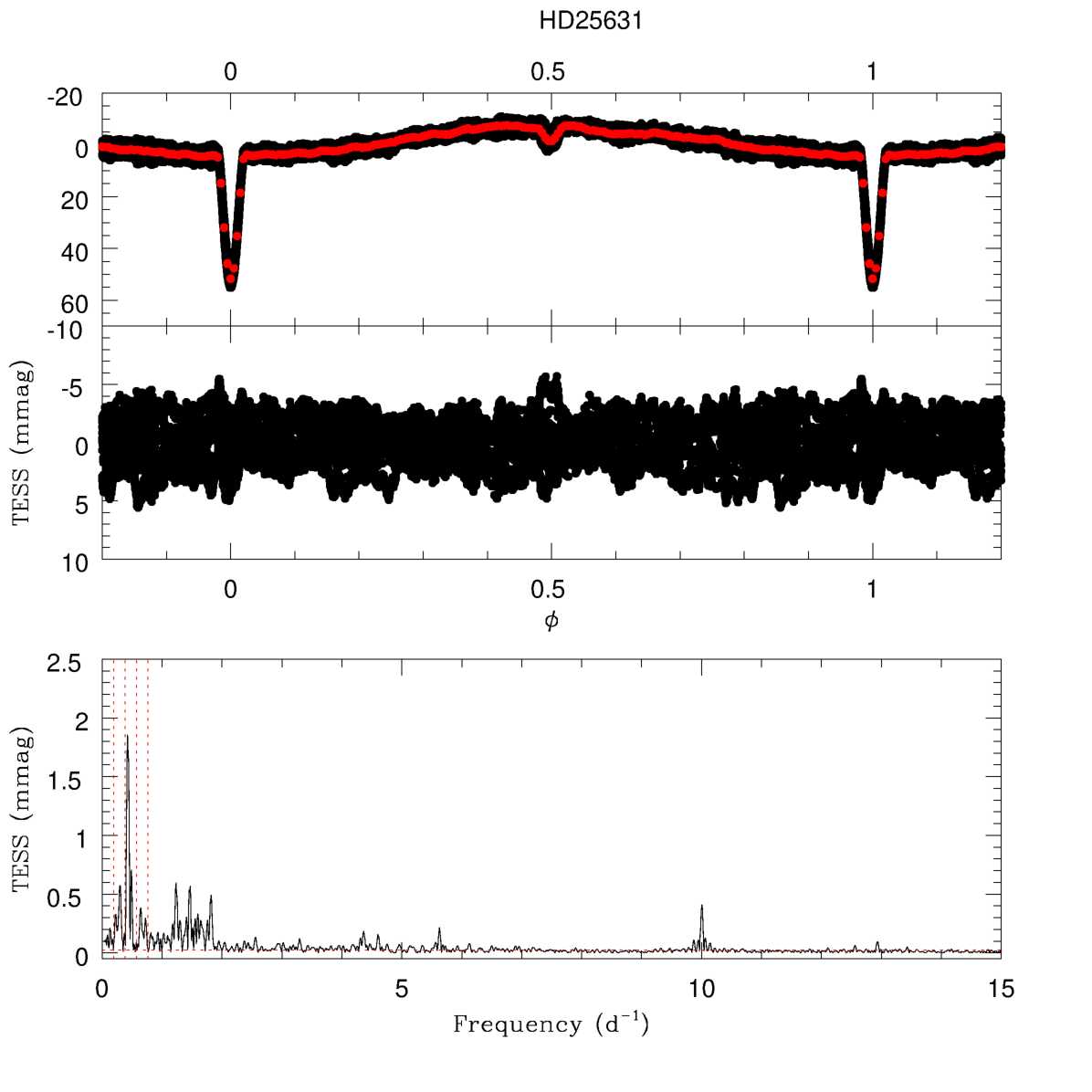}
  \end{center}
  \caption{On top, the \te\ light curve of HD\,25631, folded on the best-fit ephemeris, along with its average (in red). In the middle panel, the light curves after the average has been subtracted. At the bottom, the periodogram of that residual light curve, with the first four harmonics of the orbital period shown as vertical red dotted lines (the horizontal line corresponds to five times the noise at high frequencies).}
\label{25631cl}
\end{figure}

\begin{figure}
  \begin{center}
    \includegraphics[width=8cm]{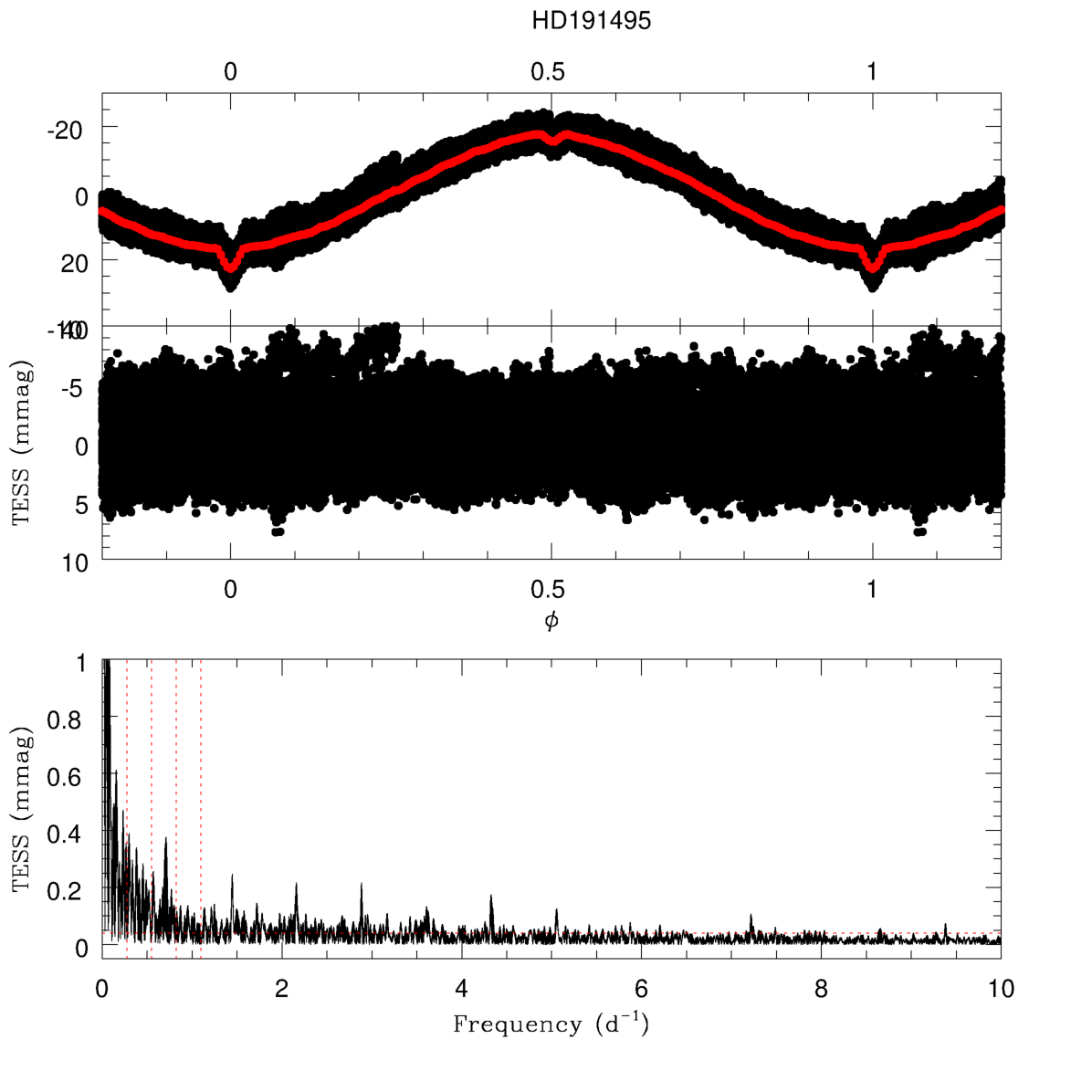}
  \end{center}
  \caption{Same as Fig. \ref{25631cl} but for HD\,191495. }
\label{191495cl}
\end{figure}

\begin{figure}
  \begin{center}
    \includegraphics[width=8cm]{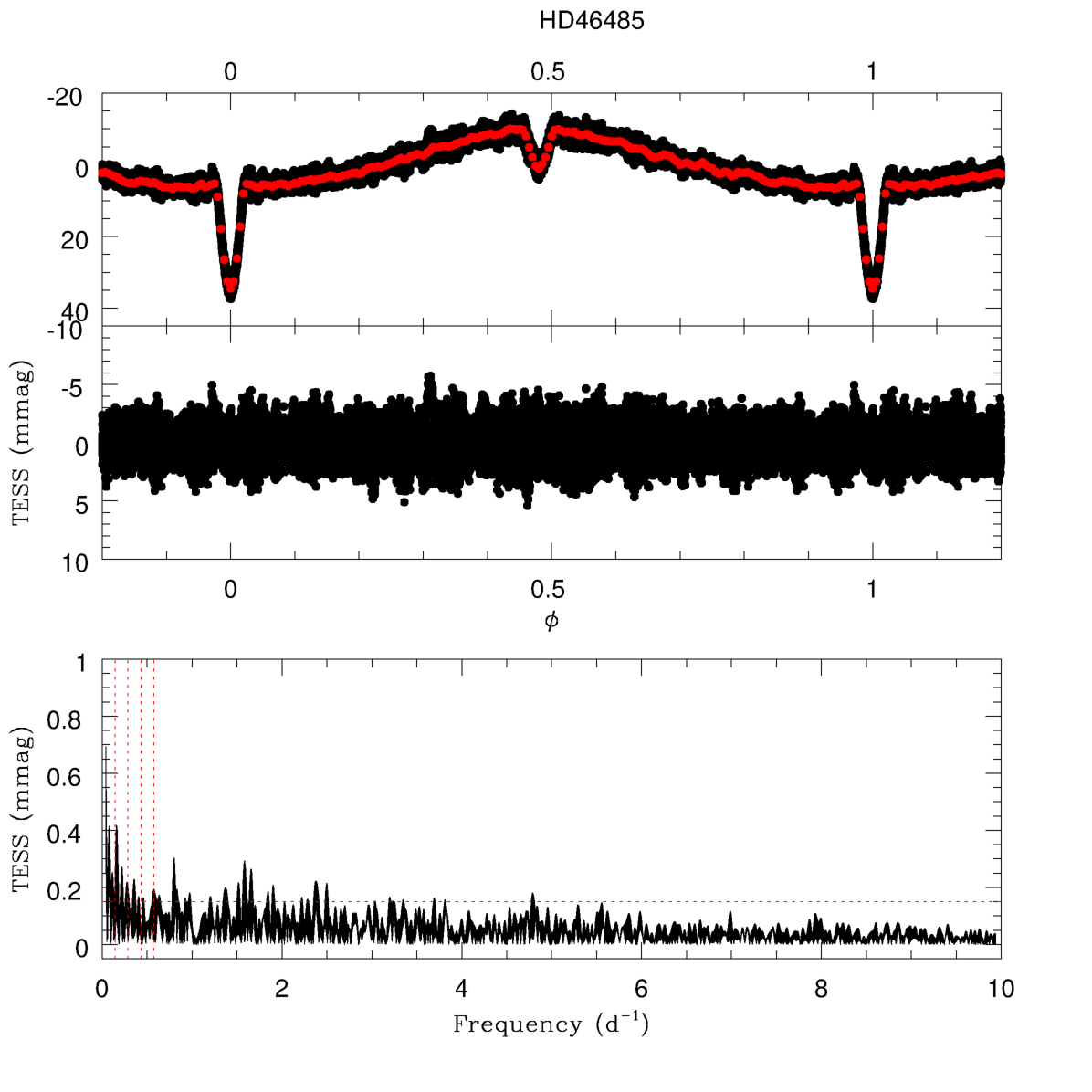}
  \end{center}
  \caption{Same as Fig. \ref{25631cl} but for HD\,46485.}
\label{46485cl}
\end{figure}

\subsubsection{Remaining photometric variability}

The average light curves were taken out from the \te\ photometry, so that only the stellar variability remains (top panels of Figures \ref{25631cl},  \ref{191495cl}, and  \ref{46485cl}). A period search was then done on this residual light curve (bottom panels of the same figures).

For HD\,25631, several peaks are detected, with amplitudes up to 2\,mmag. Their frequencies are: 0.424\,d$^{-1}$, 1.232\,d$^{-1}$, 1.468\,d$^{-1}$, 1.820\,d$^{-1}$, and 10.008\,d$^{-1}$ (small changes on the last decimal are found if restricting to data from Sector\,31, potentially indicating some variability). They are clearly not located at harmonics of the orbital frequency so they are not leftovers of the orbital signal. We may however note that the sum of 1.232 and 1.820\,d$^{-1}$ signals correspond to 16 times the orbital frequency: this could indicate that these two ``daughter'' modes come from an unstable, tidally-induced, harmonic ``parent'' \citep{guo21}. The rotation frequency of the star, derived from the observed $v\sin(i)$ after considering the best-fit inclinations and radii, amounts to $\sim$1.06\,d$^{-1}$. None of the detected peaks appears close to that frequency or its harmonics, hence they most probably reflect a true hybrid pulsational activity of the star mixing low-frequency g-modes and high-frequency p-mode. This is in line with the position of the star in the HR diagram as it indeed appears where the $\beta$\,Cep and SPB instability zones overlap \citep{mig07}.

For HD\,191495, the amplitudes at low frequencies reflect remaining long-term variations. Indeed, cycle-to-cycle variations are clearly seen when superimposing different sectors, or even during one sector. In addition, several isolated peaks are detected, with amplitudes up to 0.4\,mmag and main frequencies of 0.7104\,d$^{-1}$, 1.4466\,d$^{-1}$, 2.1598\,d$^{-1}$, 2.8842\,d$^{-1}$, 4.3252\,d$^{-1}$, and 5.0538\,d$^{-1}$. Since there are gaps between Sectors, each ``peak'' is actually formed by many densely-spaced and narrow peaks (whose width is set by the total duration of observations) within a broader modulation (whose width reflects the individual Sectors' duration). It is thus difficult to exactly pinpoint the actual frequencies. Nevertheless, it is obvious that these signals are not harmonics of the orbital frequency, but of a signal at $\sim0.72$\,d$^{-1}$. This is reminiscent of the rotational frequency. Indeed, from the observed $v\sin(i)$ and inclination derived before, that frequency corresponds to a primary radius of 5.9\,R$_{\odot}$, which agrees reasonably well with that found in the photometric solution, considering uncertainties\footnote{Even the $v\sin(i)$ is subject to non-negligible uncertainties since, for fast rotators, it strongly depends on the line examined for its derivation because of gravity darkening.}. Therefore the intrinsic variability of the \te\ light curves most probably reflects a rotational modulation which could be due to spots on the B-type star.

For HD\,46485, the residual light curve display variability with low amplitudes and, accordingly, the periodogram appears quite empty. Some stochastic red noise can be seen at low frequencies, as usual for massive stars \citep{blo11,bow20,naz21,rau21}. A few isolated peaks also exist, but their amplitudes do not exceed 0.3\,mmag. Their frequencies are 0.7998, 1.5836, 1.6575, 2.3796, and 2.4998\,d$^{-1}$ for the main ones. They do not correspond to the orbital frequency but three are harmonics of 0.8\,d$^{-1}$. Such a frequency would correspond to a rotational frequency for $R=8.6$\,R$_{\odot}$, which is reasonably close to the primary's radius. A rotational signal thus seems present in the photometry, as first suggested by \citet{bur20}. A pulsational origin is favored for the other frequencies, in line with the report by \citet{mah09} of line profile variations, which we also detect in our spectra. 

\section{Discussion}
\subsection{Other systems with similar light curves}
 The very typical shape of the light curves of our targets, which mix reflection effects and narrow grazing eclipses, has been detected before, although in a different context. In a seminal paper, \citet{moe15} examined short-period (3--8.5\,d) eclipsing systems associated to B-stars of the Large Magellanic Cloud using OGLE data. They found 22 systems exhibiting significant reflection and narrow eclipses of unequal depths, as for our targets. While no velocity study was performed for any of the targets, the light curves were fitted {\it assuming} the B-stars to be the hotter components of these systems. The remaining stellar parameters were then derived using evolutionary tracks. The massive primaries have 5--16\,M$_{\odot}$ and the derived mass ratios $q$ cover the interval 0.07--0.36. For eighteen of these systems, the companions have not yet reached the main sequence; three have companions lying on the ZAMS and the last system is contaminated by a third light hence its parameters are less well constrained. The 18 early-B+PMS systems were then named ``nascent eclipsing binaries with extreme mass ratios''. \citet{moe15} deduced from their detections, after correcting for biases, that 2\% of the early-type MS B-stars should have such low-mass, PMS companions in short-period orbits. This would imply that companions with such extreme mass ratios are ten times more frequent for B-stars than for solar-type stars. In addition, such extreme systems complement the statistical knowledge on binaries with B-type primaries: short-period systems should in fact affect $\sim$6\% of all early-type MS B-stars, one third having $q<0.25$ and two thirds having $q>0.25$ \citep{moe15}. Extreme mass ratios are thus quite common amongst close binaries, although their detection clearly remains a challenge in view of the luminosity contrast between stars. 

\citet{gul16} also searched for very low-mass companions to massive stars, but with another technique. They subtracted a theoretical spectrum from observed data of massive A- and B-stars, and then correlated the residuals with a range of model spectra. No light curve was examined; no full orbital solution could be derived as the spectroscopic analysis was only applied to one or a few spectra, but at least an estimate of the companion properties could be made. In the sample, eight cases are associated to early-type (B1--3 IV or V) stars: HD\,21428, HD\,64802, HD\,65460, HD\,133955, HD\,144294, HD\,145482, e01\,Car, and $\beta^1$\,Sco. Their companions have masses estimated to 1--4\,M$_{\odot}$, yielding mass ratios $q$ of 0.1--0.6. The distribution of the mass ratios for the whole sample peaks at 0.3 and appears quite different from the distribution observed in wide binaries, suggesting different formation paths. 

In addition to these large sample papers, a few studies of individual cases are worth mentioning. In their KELT study of $\beta$\,Cep stars, \citet{lab20} mentioned the detection of reflection effects and narrow eclipses for the {\it evolved} stars HD\,339003 (B0.5III) and HD\,254346 (B2/3III, although contamination by neighbouring stars made conclusions more difficult to reach for that latter star using only KELT data). No full light curve fitting nor velocity analysis was presented, however. \citet{ste20} analyzed the case of HD\,58730, whose primary has a later spectral type, B9V, than our targets but shares with them a rather fast rotation ($v\sin(i)=$183\,\kms). Reflection effects are not detected in the light curve of such a late-type B-star, but both ellipsoidal variations and narrow eclipses are present in its 3.6\,d cycle. Combining photometric and spectroscopic information, the secondary star was found to be a low-mass PMS star, with a mass ratio of only 0.06 and a radius inflated beyond expectations from irradiation effects. \citet{jer21} reported the presence of reflection effects, but this time without eclipses, for both $\nu$\,Cen (B2IV) and $\gamma$\,Lup\,A (B5IV). The combined analysis of their photometric and spectroscopic data led to a complete characterization of these systems, revealing PMS companions with mass ratios of 0.1--0.2. It may be noted in addition that, for $\gamma$\,Lup\,A, the primary B-star displays a fast rotation ($v\sin(i)=$236\,\kms) and that the orbit is slightly elliptical ($e=0.1$). Moreover, beyond the binary effects, the light curves also revealed $\beta$\,Cep pulsations for $\nu$\,Cen and a SPB nature for $\gamma$\,Lup\,A. Because of the presence of PMS companions, these systems were compared to the detections of \citet{moe15}, as well as the cases of $\mu$\,Eri (B5IV, \citealt{jer13}) and 16\,Lac (B2IV, \citealt{jer15}) although the light curves of the latter two stars somewhat differ. For example, their photometry displays no obvious reflection effects, longer orbital periods, a flat-bottom primary eclipse or no clear secondary eclipse. Finally, \citet{sta21} and \citet{joh21} performed a detailed (photometric+spectroscopic) analysis of the behaviour of HD\,149834 and HD\,165246, respectively, whose photometry displays both reflection effects and narrow eclipses. HD\,149834, a system belonging to NGC\,6193, appears composed of a fast-rotating B2V star ($v\sin(i)=$216\,\kms), which displays hybrid SPB/$\beta$\,Cep pulsations, and a low-mass PMS companion ($q=0.09$) in a 4.6\,d orbit. The inner binary in HD\,165246 has a similar period, but is made of a fast-rotating O8V star ($v\sin(i)=$243\,\kms) and a low-mass companion ($q=0.16$) which has not yet reached the main sequence.

\begin{figure*}
  \begin{center}
    \includegraphics[width=8cm]{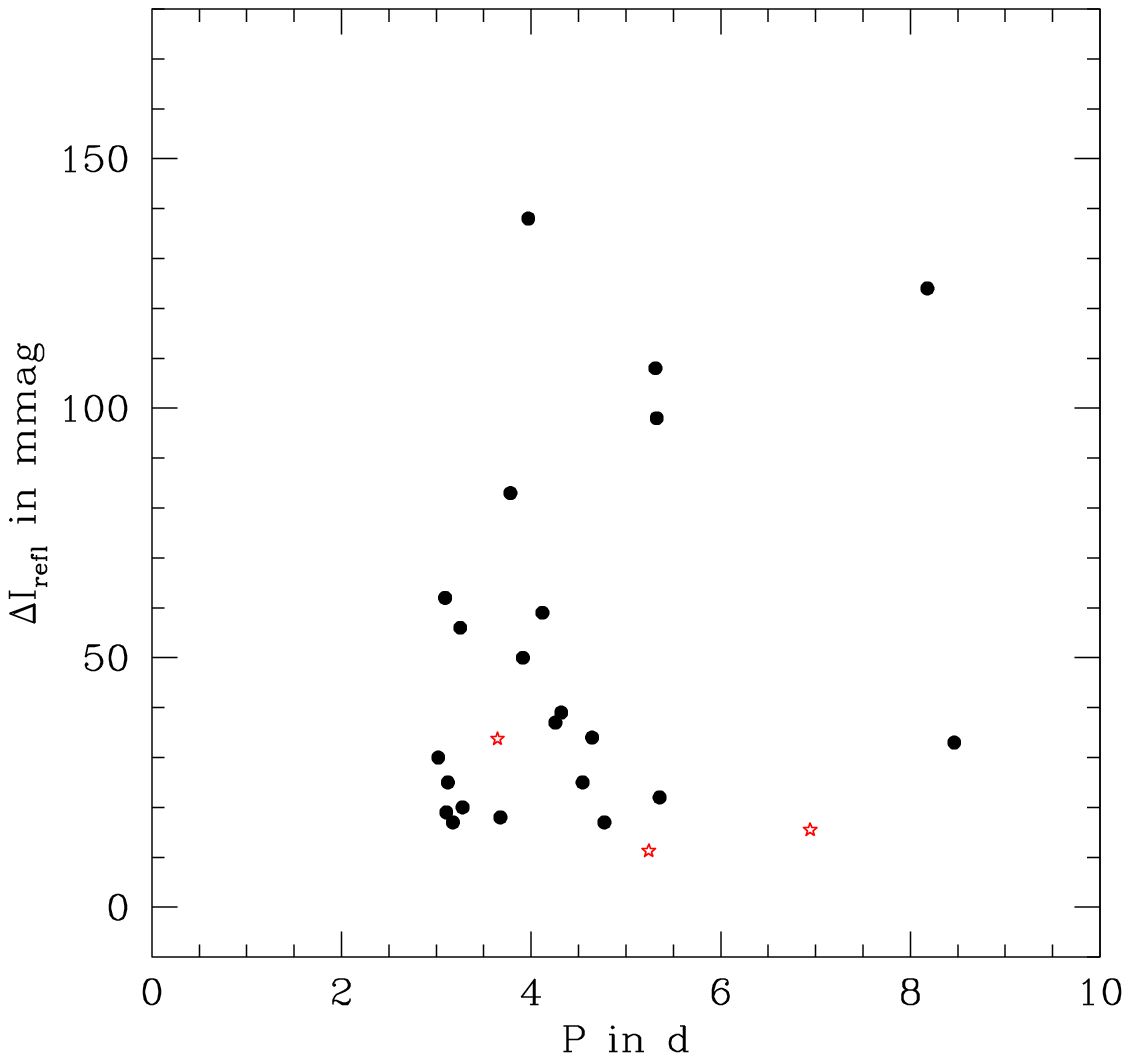}
    \includegraphics[width=8cm]{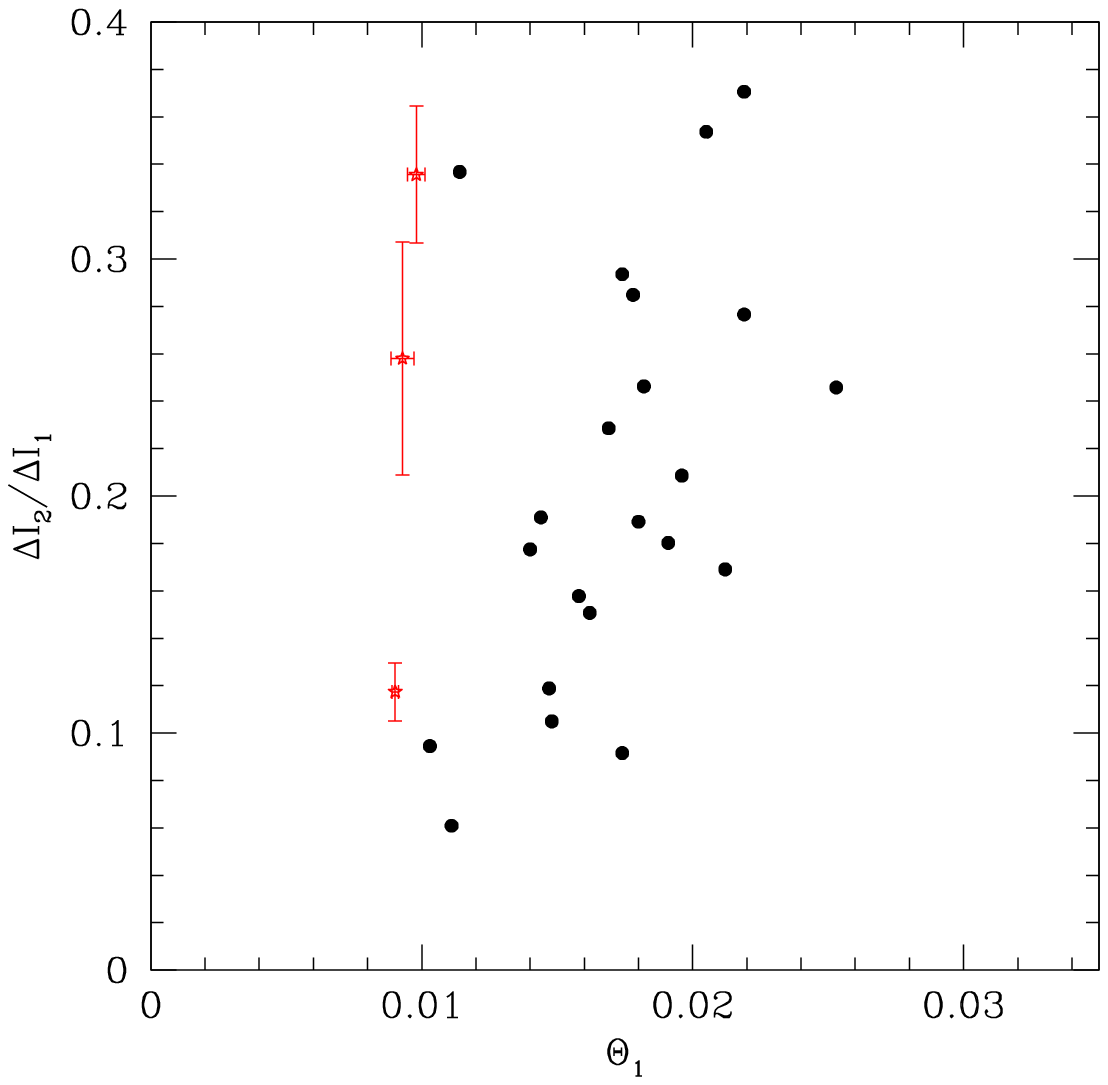}
  \end{center}
  \caption{Comparison of the light curve properties (reflection effect amplitude $\Delta I_{refl}$ and period $P$ on the left, ratio of eclipse depths $\Delta I_i$ and primary eclipse width $\Theta_1$ on the right) for the stars in this paper (red stars) with those of stars in \citet{moe15} shown as black dots.}
\label{param}
\end{figure*}

\subsection{The nature of the companions}
In this study, we have analyzed both photometry and spectroscopy of three fast-rotating main-sequence OB stars. In terms of light curves (Fig. \ref{param}, Table \ref{moefit}), our targets reveal similar properties as HD\,149834: reflection effects of 10--30\,mmag (as also found in \citealt{jer21}) and narrow eclipses (each affecting less than 5\% of the cycle). Most LMC cases of \citet{moe15} display larger reflection effects and broader eclipses, and the detection of smaller photometric changes is easily explained by the higher sensitivity of \te\ compared to ground-based observations.

Are the companions of our targets of similar nature as advocated in those previous papers? To answer this question, we put our results in two different contexts: a post-interaction scenario and a pre-interaction scenario.

  As mentioned in the introduction, fast rotation of massive stars is now often considered as the result of binary interactions. In such cases, the binaries undergo a mass-transfer event, which left the donor star stripped. The short periods of our systems (3.6--6.9\,d) may be compatible with this idea (see e.g. the work of \citealt{dem13} and the groups of short-period and long-period massive systems, aka channels 2 and 5 in \citealt{wil04}).

  Observations of systems at the different steps of such interactions are known. There are at least a dozen cases of Be stars paired with cool giants of spectral type A--K \citep{auf94,eli97,har15}. Such systems are interpreted as binaries currently undergoing mass-transfer. Their periods are found between a few days and more than 200\,d, and the masses of the cooler companions are low (0.1--5\,M$_{\odot}$). However, the companions' radii are also very large (6 to more than 100\,R$_{\odot}$) and their lines are clearly detected in the observed spectra: clearly, this does not correspond to our targets' properties. A last case is HD\,15124. The properties of its components were derived from simple SED fitting \citep{elb22} and appear rather similar to those of our targets, although the fast-rotating B-star belongs to the Be category, due to the presence of emissions linked to current accretion. However, for that system as for other post-interacting cases, the age is large: at least 200\,Myr from evolutionary models, while our targets are young ($<$20\,Myr).

  Immediately after the end of mass-transfer are found another set of systems. The spectra of both binaries HR\,6819 \citep{bod20} and ALS\,8775 (LB-1, \citealt{she20}) show narrow lines from a stripped star with moderate temperature (13--16\,kK), low mass (0.5--1.5\,M$_{\odot}$), and bloated radius (4.4--5.4\,R$_{\odot}$). The companion appears to be a fast rotating star with early B spectral type. Despite a comparable optical brightness, its lines were more difficult to spot in the spectra because of their width. A similar situation is found for NGC1850\,BH1: its spectra reveal lines of a stripped, 1--2\,M$_{\odot}$ star, whose positions change with $K\sim175$\,\kms\ during the 5\,d orbit \citep{sar23}. The more massive, $\sim$5\,M$_{\odot}$, companion is faint in the optical and its signature could not yet be isolated. Such cases appear very different from our targets (which are not known to be or have been Be stars, although the presence of a disk may be a transient feature). Indeed, the spectra clearly show first and foremost the lines of the more massive, fast-rotating B-type star. Also, while masses and radii of the stripped stars are rather comparable to those found for the companions in this work, their temperatures (13--16\,kK) appear on the highest side of those derived here.  

At the next evolutionary step, the stripped star is expected to settle into a hot subdwarf with sdOB type. More than a dozen such companions to fast-rotating Be stars were detected through their UV emissions \citep{wan21}. Their small size makes them good candidates to produce narrow eclipses, but the detailed optical and UV analysis of our targets (see above) demonstrates that such hot objects are not present in our targets.

As a last check, we searched the database of Binary Population And Spectral Synthesis v2.2.1 binary models\footnote{https://bpass.auckland.ac.nz/} (BPASS, \citealt{eld17,ste20b} and references therein) for post-interactions systems with solar metallicity ($Z=0.02$) and parameters similar to those of our three systems. The first screening criteria were period and masses\footnote{Note that the mass gainer is star \#2 in BPASS but is called ``primary'' in this paper, and the mass donor is star \#1 in BPASS but called ``secondary'' in this paper.}. Adequate models could be found only for HD\,25631. They were further filtered to find amongst them those also reproducing the observed stellar radii: only two models fulfilled these criteria. They have initial masses of 7.5 or 8.0\,M$_{\odot}$ for the mass donor and 0.8\,M$_{\odot}$ for the mass gainer. The evolution stage at which period, masses, and radii observed for HD\,25631 are reproduced in models corresponds to a mass donor being a stripped star with temperature around 40\,000\,K, although no such hot companion is seen in our data. Furthermore, that modelled system has an age near 40\,Myr, which is well above that derived for HD\,25631. 

In summary, there seems to be little support for a post-interaction scenario for our targets. Let us now examine the pre-interaction scenario. The masses and temperatures derived for the companions in our systems are compatible with low-mass stars, which would then still be in their PMS stage in view of the systems' ages and bloated radii of the companions. That interpretation was previously favored by several authors for objects presenting light curves and RV curves similar to those of our targets \citep{moe15,ste20,jer21,sta21,joh21}. 

In addition, the most likely explanation for the properties of the reconstructed secondary spectrum of HD\,25631 is that we are dealing with a chromospherically active pre-main sequence or zero-age main-sequence star. \citet{Yam20} present observations of the Ca\,{\sc ii} near-IR triplet in a sample of 60 PMS stars. These objects all display strong Ca\,{\sc ii} emission lines, with a majority (53 out of 60) displaying rather narrow emissions similar to what we have found here. We measured the equivalent widths (EWs) of the Ca\,{\sc ii} emission lines on the reconstructed secondary spectrum. We obtained EW ratios of EW(Ca\,{\sc ii} $\lambda$\,8542)/EW(Ca\,{\sc ii} $\lambda$\,8498) $ = 1.5 \pm 0.3$ and EW(Ca\,{\sc ii} $\lambda$\,8662)/EW(Ca\,{\sc ii} $\lambda$\,8498) $ = 0.9 \pm 0.2$. These values are consistent with the ranges (1.0 -- 2.0 and 0.7 -- 1.7) reported by \citet{Yam20} for their sample of PMS stars. Given the similarity of these ratios with the solar value, \citet{Yam20} suggested that the Ca\,{\sc ii} emissions arise in regions that are similar to solar flares and plages. The Li\,{\sc i} $\lambda$\,6708 line could provide further hints on the youth of the star. Its EW appears to be about $100$\,m\AA. Yet, this EW value must be taken with caution since the disentangled spectrum is a mix of spectra of hemispheres that are most probably at very different temperatures.

\begin{figure}
  \begin{center}
    \includegraphics[width=8cm]{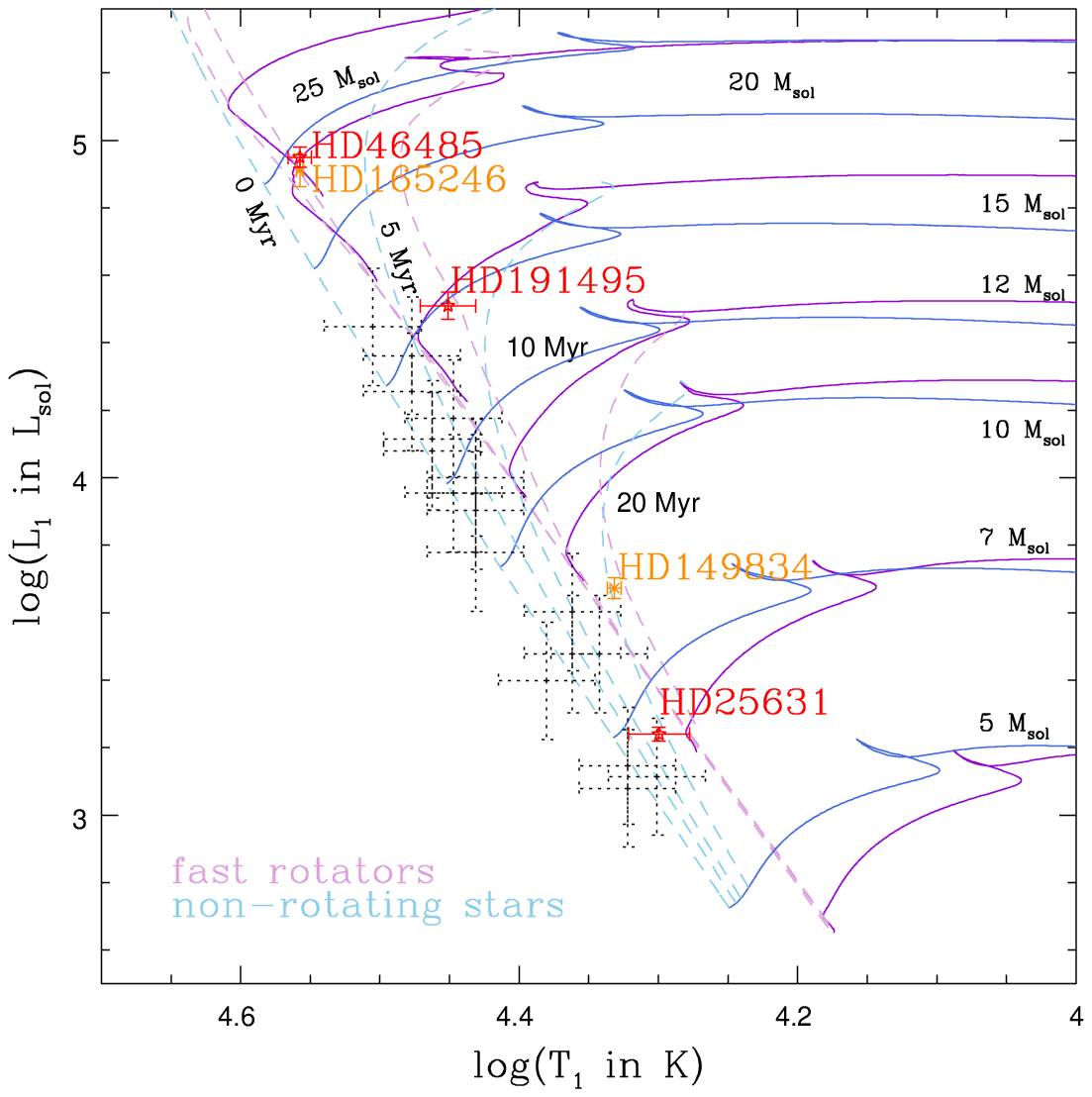}
    \includegraphics[width=8cm]{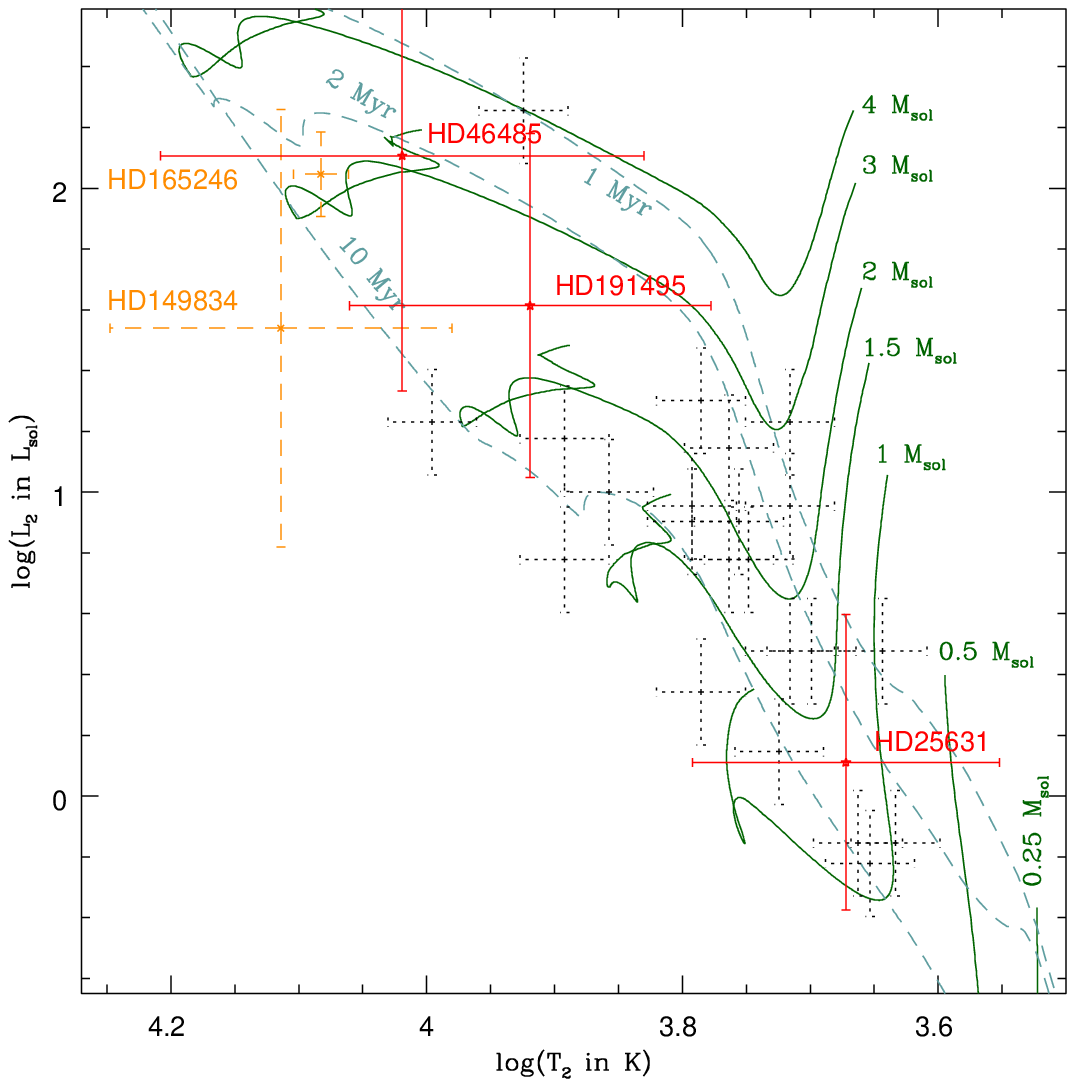}
  \end{center}
  \caption{Hertzsprung-Russell diagrams with the post-PMS evolutionary tracks of \citet{bro11} with Milky Way abundances (top, for the massive primaries) and the PMS tracks of \citet{tog11} for standard solar parameters (bottom, for the low-mass secondaries). The stars from \citet{moe15}, \citet{sta21}+\citet{joh21}, and this work are shown with black crosses with dotted lines, orange crosses with dashed lines, and red dots with solid lines, respectively. Note that the \citet{moe15} objects belong to the LMC hence their metallicity is not solar and that, in top panel, tracks from \citet{bro11} are both shown without initial rotation (blue lines) and with the fastest initial rotation available in these models (540--595\,\kms, depending on mass, violet lines). Isochrones are shown in lighter color and with dashed lines.}
\label{hrd1}
\end{figure}

\begin{figure}
  \begin{center}
    \includegraphics[width=8cm]{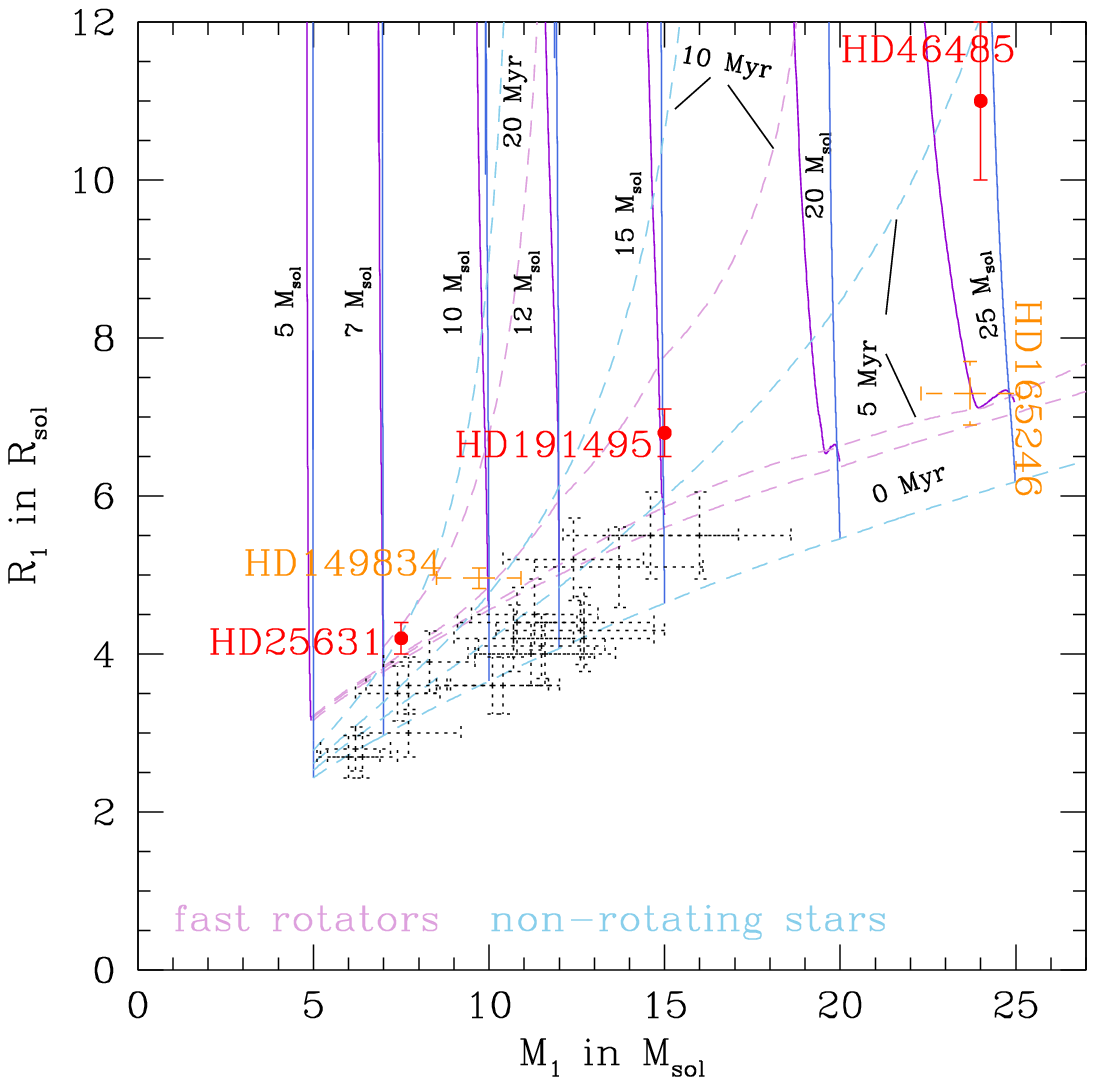}
    \includegraphics[width=8cm]{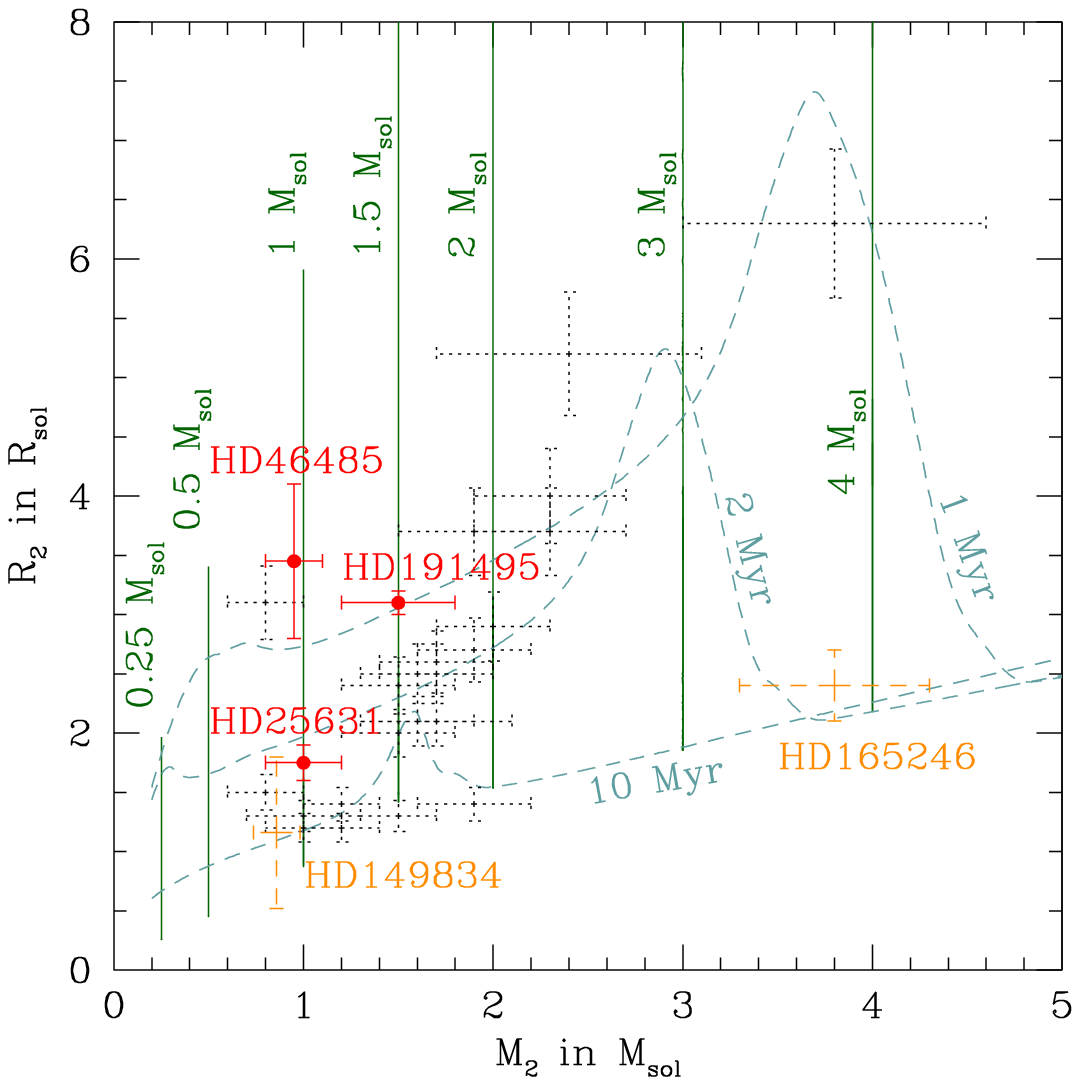}
  \end{center}
  \caption{Same as Fig. \ref{hrd1} but for radius-mass diagrams.}
\label{hrd2}
\end{figure}

The properties of our targets can then be compared to appropriate evolutionary models \citep{bro11,tog11}. In the Hertzprung-Russell diagram (Fig. \ref{hrd1}), our massive OB-stars lie amongst the primaries of the systems studied by \citet{moe15}, \citet{sta21}, and \citet{joh21}. For the cooler companions, the uncertainty in temperature (hence in luminosity) makes more difficult to judge any agreement or misfit with tracks in the Hertzprung-Russell diagram\footnote{The masses of the secondaries in HD\,46485 and HD\,149834 appear larger (2--4\,M$_{\odot}$) if derived from the evolutionary tracks of \citet{tog11}, but the large errors prohibit any definitive conclusion.}. To avoid this uncertainty, we rather draw radius-mass diagrams in Figure \ref{hrd2}, as these values appear much more secure than the secondary temperatures. Both masses and radii of primaries and secondaries appear in line with predictions from evolutionary models. Furthermore, isochrones in Figure \ref{hrd2} indicate that the young ages of the systems are in line with the BONNSAI fitting (Table \ref{results}) and the cluster ages (Sect. 3.1). For HD\,25631, the primary's age cannot be much larger than 20\,Myr while the secondary's age is 2--10\,Myr. Note that, for HD\,149834, both ages appear to be around 10\,Myr. In HD\,46485, the massive primary has an age below 5\,Myr, and the low-mass secondary has an age $\gtrsim1$\,Myr. For HD\,191495, the primary's position indicates an age below 10\,Myr (most probably 5--10\,Myr) while the secondary corresponds to a somewhat smaller age of $\sim1$\,Myr. Considering the uncertainties and the fact that models correspond to single stars (without peculiar effects such as one-sided illumination), all this appears as a very good agreement. Our targets thus seem to be new examples of ``nascent eclipsing binaries with extreme mass ratios''. It may be remarked at this stage that, for such low mass ratios, the synchronization of the stars' rotation will take much longer than for cases with more massive companions hence fast rotation will survive longer. In retrospect, young fast rotators thus appear as the best candidates to search for such extreme mass-ratio systems.

One may wonder how such close pairs with extreme ratios are formed. Their formation is often regarded as differing from that of wide binaries or from that of equal-mass systems \citep{gul16}. The main scenario considers a fragmentation of the protostellar disk \citep{kra06}. For massive protostars, the radiative feedback is known to inhibit such fragmentation, but the episodic nature of the accretion allows the formation of low-mass companions \citep{sta11}. Further accretion onto the fragments could lead to a subsequent trend towards mass equalization, but this is countered by the action of magnetic fields which favor accretion onto the massive component \citep{zha13}. In this context, the fragments can be at first more widely separated, but then interact and migrate towards each other thereby reaching the current close configuration \citep{kra06}. An alternative scenario considers that the stars are born separately but dynamic interactions, especially those involving a third star, could lead to the formation of a close unequal pair \citep{moe07,nao14}. A combination of these scenarios is also possible.

Regarding the fate of such PMS+OB-star pairs, it is expected that the systems will rapidly change as the most massive star evolves fast. Due to their proximity, it is then probable that the stars will undergo a common envelope phase - preliminary investigation with MESA suggests this will happen within 10\,Myrs. The massive star may then merge with its companion, as is often predicted for such extreme mass ratio cases (\citealt{van97} consider it as the most likely outcome for all systems with $q<0.2$, \citealt{wel01} for all systems with $q<0.65$ at the onset of the mass-transfer). Alternatively, if both objects survive, the massive star would leave behind a compact remnant (white dwarf, neutron star). After some time, the companion will finally evolve too. The system will then give birth to a low-mass X-ray binary (possibly harbouring a ms-pulsar) or to a double-degenerate WD pair which will end in a SN type Ia event. To be more definitive on the final outcomes of our systems, the impact of all physical processes (e.g. stellar winds, non-conservative mass transfer, see e.g. \citealt{men21}) needs to be properly investigated, which is beyond the scope of the present paper (but will be adressed in Britavskiy et al., in prep). It is to be hoped that more examples of such extreme mass-ratio binaries will be found and fully characterized in the future. This will help assessing their incidence amongst the populations of massive stars, notably for population synthesis calculations.

\section{Summary and Conclusion}

In this paper, we have investigated the multiplicity of three massive fast rotators in detail: HD\,25631, HD\,191495, and HD\,46485. Their \te\ light curves display a 10--30\,mmag reflection effect and two narrow eclipses of 2--50\,mmag amplitudes. The combined analysis of photometry and spectroscopy indicates that the massive stars actually are the hottest components of these systems. The companions display typical masses around 1\,M$_{\odot}$ and radii 2--4\,R$_{\odot}$. While the secondary temperatures appear somewhat uncertain, inclination, mass ratio, and radii appear quite robust. The absence of UV signatures from the companions suggests that they are not hot subdwarfs and the observed data also appear different from those of systems with recent mass-transfer. As was done for similar systems (e.g. HD\,149834, HD\,165246, two dozen cases in the LMC), an interpretation in terms of pre-interacting OB+PMS systems thus appears probable. This is notably corroborated by the detection of near-IR Ca\,{\sc ii} lines in emission for HD\,25631 and the young ages of the systems. The fast rotation of the massive stars would then be primordial for those objects. 

\section*{Acknowledgements}
Y.N., N.B., and G.R. acknowledge support from the Li\`ege University (IPD-STEMA funding), the Fonds National de la Recherche Scientifique (Belgium), the European Space Agency (ESA) and the Belgian Federal Science Policy Office (BELSPO) in the framework of the PRODEX Programme (contracts linked to XMM-Newton). S.S-D acknowledges support from the Spanish Ministry of Science and Innovation (MICINN) through the Spanish State Research Agency through grants PGC-2018-0913741-B-C22, PID2021-122397NB-C21, and the Severo Ochoa Programe 2020-2023 (CEX2019-000920-S). This work has also received financial support from the Canarian Agency for Economy, Knowledge, and Employment and the European Regional Development Fund (ERDF/EU), under grant with reference ProID2020010016. ADS and CDS were used for preparing this document. 

\section*{Data Availability}
The \te, Espadons, IUE, FUSE, and ESO data presented in this article are available in their respective public archives. The TIGRE data are available upon reasonable request.

\bsp	
\label{lastpage}
\end{document}